\pdfoutput=1
\newcommand*{\ATLASLATEXPATH}{latex/}
\documentclass[cernpreprint, texlive=2016, UKenglish]{\ATLASLATEXPATH atlasdoc}
\usepackage{multirow}
\usepackage{comment}
\usepackage{rotating}
\usepackage{pdflscape}
\usepackage{\ATLASLATEXPATH atlaspackage}
\usepackage{\ATLASLATEXPATH atlasbiblatex}
\usepackage{\ATLASLATEXPATH atlasphysics}
\graphicspath{{logos/}{figures/}}

\usepackage{EXOT-2017-25-PAPER-defs}

\AtlasTitle{Search for long-lived neutral particles in $pp$ collisions at ${\sqrt{s}} = 13~\TeV$ that decay into displaced hadronic jets in the ATLAS calorimeter}

\AtlasAbstract{
 
This paper describes a search for pairs of neutral, long-lived particles decaying in the ATLAS calorimeter.
Long-lived particles occur in many extensions to the Standard Model and may elude searches for new promptly decaying particles.
The analysis considers neutral, long-lived scalars with masses between $5\GeV$ and $400\GeV$, 
produced from decays of heavy bosons with masses between $\mPhiLo \GeV$ and $\mPhiHi \GeV$, where the long-lived scalars decay into Standard Model fermions.
The analysis uses either \intlumiNOMATCH or \intlumi of data (depending on the trigger) recorded in 2016 at the LHC with the ATLAS detector in proton--proton collisions at a centre-of-mass energy of 13~\TeV.
No significant excess is observed, and limits are reported on the production cross section times branching ratio as a function of the proper decay length of the long-lived particles.
}

\AtlasJournal{EPJC}
\PreprintIdNumber{CERN-EP-2018-351}

\AtlasJournalRef{Eur. Phys. J. C 79 (2019) 481}
\AtlasDOI{10.1140/epjc/s10052-019-6962-6}

\hypersetup{pdftitle={ATLAS document},pdfauthor={The ATLAS Collaboration}}

\begin{document}

\maketitle

\newpage

\section{Introduction}
\label{sec:intro}
Long-lived particles (LLPs) feature in a variety of models that have been proposed to address some of the open questions of the Standard Model (SM). Examples are:
various supersymmetric (SUSY) models~\cite{Arvanitaki:2012ps,ArkaniHamed:2012gw,Giudice:1998bp,Barbier:2004ez,Csaki:2015fea,Fan:2011yu,Fan:2012jf}; Neutral Naturalness~\cite{Chacko:2015fbc,Burdman:2006tz,Cai:2008au,Chacko:2005pe} and Hidden Sector (HS)~\cite{Strassler:2006im,Strassler:951513,Chan:2011aa} models that address the hierachy problem; models that seek to incorporate dark matter~\cite{Baumgart:2009tn,Kaplan:2009ag,Dienes:2011ja,Dienes:2012yz}, or explain the matter–antimatter asymmetry of the universe~\cite{Cui:2014twa}; and models that lead to massive neutrinos~\cite{Helo:2013esa,Batell:2016zod}.
Decays of LLPs created in collider experiments would produce unique signatures that may have been overlooked by previous searches for particles that decay promptly. This paper presents a search sensitive to neutral LLPs decaying mainly in the hadronic calorimeter (HCal)
or at the outer edge of the electromagnetic calorimeter (ECal) of the ATLAS detector. This allows the analysis to probe LLP proper decay lengths ($c\tau$, where $c$ is the speed of light and $\tau$ is the lifetime of the LLP) ranging between a few centimetres and a few tens of metres.
In HS models, a proposed new set of particles and forces is weakly coupled to the SM via a mediator particle.
As a benchmark, this analysis uses a simplified HS model~\cite{Chung2006,Chang2008,Strassler:2006im,Strassler:951513,Chan:2011aa}, in which the SM and HS are connected via a heavy neutral boson ($\Phi$), which may decay into two long-lived neutral scalar bosons ($s$). The neutral scalars are assumed not to interact with the detector.
While $\Phi$ could be the Higgs boson, this analysis considers mediators with masses ranging from $\mPhiLo \GeV$ to $\mPhiHi \GeV$, and scalars with masses between $\mSLo \GeV$ and $\mSHi \GeV$.
The decay \higgsd is considered, where $f$ refers to fermions.
Decays to bosons are not considered in the benchmark model used in this analysis.
Since this model assumes that the branching ratios of the scalar decaying into SM fermions are the same as those of the SM Higgs, each long-lived scalar usually decays into heavy fermions: $b\bar{b}$, $c\bar{c}$, and $\tau^{+}\tau^{-}$.
The branching ratio among the different decays depends on the mass of the scalar but for $\mS \geq 25~\GeV$ it is almost constant and equal to 85:5:8.
The SM quarks from the LLP decay hadronize, resulting in jets whose origins may be far from the interaction point (IP) of the collision. 
The proper decay lengths of LLPs in HS models are typically unconstrained, aside from a rough upper limit of $c\tau \lesssim 10^8$~m given by the cosmological constraint of Big Bang Nucleosynthesis~\cite{Jedamzik:2006xz}, and could be short enough for the LLPs to decay inside the ATLAS detector volume. 

Previous searches for pair-produced neutral LLPs at hadron colliders have been performed at the Tevatron and at the LHC.
At the Tevatron, searches by D$\emptyset$~\cite{HVD0} and CDF~\cite{HVCDF} looked for displaced vertices in their tracking system only, allowing them to set limits on LLP proper decay lengths of the order of a few centimetres.
At the LHC, the CMS experiment has performed searches at centre-of-mass energies of 7, 8 or 13~\tev for neutral LLPs by considering events with either converted photons and missing energy \cite{CMS-EXO-11-035,CMS-EXO-11-067}, or with lepton~\cite{CMS-EXO-11-101,CMS-EXO-12-037} or jet pairs~\cite{Sirunyan:2018pwn, CMS-EXO-18-007} originating from displaced vertices in the tracking system. A CMS search for jet pairs originating in the tracker was also performed at 13~\tev~\cite{CMS-EXO-16-003}. The CMS searches are sensitive to LLP proper decay lengths from $\sim 0.1$~mm to $\sim 2$~m.
Previous ATLAS searches for neutral LLPs consider events with photons~\cite{SUSY-2013-17}, or particles originating from displaced vertices in the tracking system~\cite{Aaboud:2017iio,Aad:2015rba}. Other searches involve pairs of displaced jets in the HCal (8~\tev)~\cite{EXOT-2012-28,Aaboud:2018arf}, or pairs of reconstructed vertices in the muon spectrometer (MS) at 7 and 13~\tev~\cite{EXOT-2011-11,Aaboud:2018aqj}, or the combination of one displaced vertex in the MS and one in the inner tracking detector (8~\tev)~\cite{EXOT-2013-12}. Other searches consider pairs of muons originating after the inner tracker~\cite{Aaboud:2018jbr,Aad:2014yea}.
These ATLAS searches are complementary, since they use different sub-detectors, and therefore their sensitivities are governed by different instrumental effects and sub-detector responses to the kinematics of the LLP decays. They also have different backgrounds, and different lifetime coverage due to the different physical location of the sub-detectors, with sensitivity to LLP proper decay lengths extending from a few millimeters to about $200$~m.

The analysis presented in this paper is an update to the 8~\tev ATLAS search for pair-produced neutral LLPs decaying in the HCal~\cite{EXOT-2012-28}, using \intlumiNOMATCH or \intlumi of 13~\tev data depending on the trigger, with significant improvements to the displaced-jet identification, event selection and background estimation.
If the scalar decay occurs in the calorimeters, the two resulting quarks are reconstructed as a single jet with unusual features compared to jets from SM processes.
These jets will typically have no associated activity in the tracking system. Furthermore, they will often have a high ratio of energy deposited in the HCal ($E_{\mathrm{H}}$) to energy deposited in the ECal ($E_{\mathrm{EM}}$). This ratio, \calrat, is referred to as the \emph{CalRatio}.
Finally, jets resulting from these decays will appear narrower than prompt jets when reconstructed with standard algorithms.
This analysis requires two such non-standard jets.

The main background process that mimics this signature is SM multijet production, in cases where the jets are composed mainly of neutral hadrons or are mis-reconstructed due to noise or instrumental effects.
Despite the low probability of a prompt jet to produce a signal-like jet, the SM multijet rate is high enough for this to be the dominant background.
Other contributions come from the non-collision background consisting of cosmic rays and beam-induced background (BIB)~\cite{CERN-EP-2016-029}.
The latter is composed of LHC beam--gas interactions and beam-halo interactions with the collimators upstream of the ATLAS detector, resulting in muons travelling parallel to the beam-pipe.

Two triggers were used to collect the data, one optimal for models with $\mPhi > 200\GeV$ and the other for $\mPhi \leq 200\GeV$, and different selections are used to analyse the dataset collected with each trigger.
Jets are classified as signal- or background-like jets using machine learning in two steps:
first, for every reconstructed jet, a multilayer perceptron, trained on signal jets from LLP decays, is used to predict the decay position of the particle that generated it;
next, a per-jet Boosted Decision Tree (BDT) classifies jets as signal-like, multijet-like or BIB-like jets.
Events are then classified as likely to have been produced by a signal process or a background process using a per-event BDT. Two separate versions of the per-event BDT are trained: one optimised for models with $\mPhi\leq 200\gev$ (referred to as \emph{low-}\mPhi models), and the other for models with $\mPhi> 200\gev$ (\emph{high-}\mPhi models).
The final sample is constructed by making a selection on the relevant per-event BDT output value of candidate events and imposing event quality criteria and requirements to suppress cosmic rays and BIB. These selections remove almost all the non-collision background, leaving only multijet background, and maximise signal-to-background ratio in the final search region.

The ATLAS detector is described in Section~\ref{sec:detector}. The collection of the data and generation of samples of simulated events are then discussed in Section~\ref{sec:datasamples}. The trigger and event selection are detailed in Section~\ref{sec:selection}, followed by a discussion of the estimate of the background yield in the search regions in Section~\ref{sec:background}. The systematic uncertainties are summarised in Section~\ref{sec:systematics}. The statistical interpretation of the data and combination of results with the MS displaced vertex search are described in Section~\ref{sec:interpretation}, and the conclusions are given in Section~\ref{sec:conclusion}.

\section{ATLAS detector}
\label{sec:detector}

\newcommand{\AtlasCoordFootnote}{
ATLAS uses a right-handed coordinate system with its origin at the nominal interaction point
in the centre of the detector and the $z$-axis along the beam pipe.
The $x$-axis points from the IP to the centre of the LHC ring,
and the $y$-axis points upwards.
Cylindrical coordinates $(r,\phi)$ are used in the transverse plane,
$\phi$ being the azimuthal angle around the $z$-axis.
The pseudorapidity is defined in terms of the polar angle $\theta$ as $\eta = -\ln \tan(\theta/2)$.
Angular distance is measured in units of $\Delta R \equiv \sqrt{(\Delta\eta)^{2} + (\Delta\phi)^{2}}$.}

The ATLAS detector~\cite{PERF-2007-01} at the LHC covers nearly the entire solid angle around the collision point.\footnote{\AtlasCoordFootnote}
It consists of an inner tracking detector surrounded by a thin superconducting solenoid, electromagnetic and hadronic calorimeters,
and a muon spectrometer incorporating three large superconducting toroidal magnets.
The inner-detector system is immersed in a \SI{2}{\tesla} axial magnetic field
and provides charged-particle tracking in the range $|\eta| < 2.5$.

The high-granularity silicon pixel detector covers the vertex region and typically provides four measurements per track.
The layer closest to the interaction point is known as the insertable B-layer~\cite{Capeans:1291633,CERN-LHCC-2012-009,Abbott:2018ikt}. It was added in 2014 and provides high-resolution hits at small radius to improve the tracking performance.
The pixel detector is surrounded by the silicon microstrip tracker, which usually provides four three-dimensional measurement points per track.
These silicon detectors are complemented by the transition radiation tracker, with coverage up to $|\eta| = 2.0$, which enables radially extended track reconstruction in this region.

The calorimeter system covers the pseudorapidity range $|\eta| < 4.9$.
Within the region $|\eta|< 3.2$, electromagnetic calorimetry is provided by barrel and endcap high-granularity lead/liquid-argon (LAr) electromagnetic calorimeters, with an additional thin LAr presampler covering $|\eta| < 1.8$, to correct for energy loss in material upstream of the calorimeters.
The ECal extends from $1.5$~m to $2.0$~m in radial distance $r$ in the barrel and from $3.6$~m to $4.25$~m in $|z|$ in the endcaps.
Hadronic calorimetry is provided by a steel/scintillator-tile calorimeter, segmented into three barrel structures within $|\eta| < 1.7$, and two copper/LAr hadronic endcap calorimeters covering $|\eta| > 1.5$.
The HCal covers the region from $2.25$~m to $4.25$~m in $r$ in the barrel (although the HCal active material extends only up to $3.9$~m) and from $4.3$~m to $6.05$~m in $|z|$ in the endcaps.
The solid angle coverage is completed with forward copper/LAr and tungsten/LAr calorimeter modules optimised for electromagnetic and hadronic measurements respectively.

The calorimeters have a highly granular lateral and longitudinal segmentation. Including the presamplers, there are seven sampling layers in the combined central calorimeters (the LAr presampler, three in the ECal barrel and three in the HCal barrel) and eight sampling layers in the endcap region (the presampler, three in ECal endcaps and four in HCal endcaps). The forward calorimeter modules provide three sampling layers in the forward region. 
The total amount of material in the ECal corresponds to $24 - 35$ radiation lengths in the barrel and $35 - 40$ radiation lengths in the endcaps. The combined depth of the calorimeters for hadronic energy measurements is more than 9 hadronic interaction lengths nearly everywhere across the full detector acceptance. 

The muon spectrometer comprises separate trigger and
high-precision tracking chambers measuring the deflection of muons in the magnetic field generated by the superconducting air-core toroids.
The field integral of the toroids ranges between \num{2.0} and \SI{6.0}{\tesla\metre}
across most of the detector.

The ATLAS detector selects events using a tiered trigger system~\cite{TRIG-2016-01}. The level-1 trigger is implemented in custom electronics and reduces the event rate from the LHC crossing frequency of 40~MHz to a design value of 100~kHz. The second level, known as the high-level trigger, is implemented in software running on a commodity PC farm that processes the events and reduces the rate of recorded events to 1~kHz.

\section{Data and simulation samples}
\label{sec:datasamples}

\subsection{Data samples}
\label{sec:data}

The data used in this analysis were collected by the ATLAS detector during 2016 data-taking using proton--proton ($pp$) collisions at $\sqrt{s}=13\TeV$.
Four datasets are defined according to the trigger used to select them.
The search is performed on the so-called \emph{main} dataset, collected by two different LLP signature-driven triggers, referred to as the low-\et CalRatio trigger and high-\et CalRatio trigger, which are described in detail in Section~\ref{sec:selection}. The high-\et CalRatio trigger was active during the full 2016 data-taking period. After requirements based on beam and detector conditions and data quality are applied, the data collected with this trigger corresponds to an integrated luminosity of \intlumi. The low-\et CalRatio trigger was activated in September 2016, collecting data corresponding to an integrated luminosity of \intlumiNOMATCH. The events collected with these triggers are referred to as high-\et and low-\et datasets respectively.
Two additional datasets, referred to as the BIB and cosmics datasets, were collected using dedicated triggers running in special conditions, as described in Section~\ref{sec:selection}. 

\subsection{Signal and background simulation}
\label{sec:sig_bkgMC}

The $\Phi \rightarrow s s$ signal samples were generated using \textsc{MadGraph5}~\cite{MadGraphNLO} at leading order (LO) with the \textsc{NNPDF2.3LO} parton distribution function (PDF) set~\cite{Ball:2012cx}. The shower process was implemented using \textsc{Pythia 8.210} \cite{Sjostrand:2014zea} using the \textsc{A14} set of tuned parameters (tune)~\cite{ATL-PHYS-PUB-2014-021}. 
Several sets of samples were generated, each modelling different combinations of \mPhi and \mS, with $\mPhi \in [125,1000]\gev$ and $\mS \in [5, 400]\gev$. 
For consistency with the rest of the samples, in the $\mS = 400\gev$ case, top-quark decays were not included in the generation process, even though they are kinematically allowed.
The simplified model used in the generation does not give a specific prediction for the absolute production cross section.
Each sample was generated for two assumptions about the LLP decay length: one sample is used to study the signal throughout the analysis, while the other sample (with the alternate decay length assumption) is used in the training of the BDTs as well as to validate the procedure for extrapolating limits to different proper decay lengths of the long-lived scalar $s$. 

The main SM background in this analysis is multijet production. Although a data-driven method is used to perform the background estimation, simulated multijet events are needed for BDT training and evaluation of some of the systematic uncertainties.
The samples were generated with \textsc{Pythia 8.186} \cite{Sjostrand:2007gs} using the \textsc{A14} tune for 
parton showering and hadronisation. The \textsc{ NNPDF2.3LO} PDF  set was used. 

To model the effect of multiple $pp$ interactions in the same or neighbouring bunches (pile-up), simulated inclusive $pp$ events were overlaid on each generated signal and background event. 
The multiple interactions were simulated with \textsc{Pythia 8.186} using the A2 tune~\cite{ATL-PHYS-PUB-2012-003} and the 
MSTW2008LO PDF set~\cite{Martin:2009iq}. 

The detector response to the simulated events was evaluated with the \textsc{GEANT4}-based detector simulation~\cite{SOFT-2010-01, Geant4}. 
A full simulation of all the detector components was used for all the samples. 
The standard ATLAS reconstruction software was used for both simulation and $pp$ data.

\section{Trigger and event selection}
\label{sec:selection}

Events are first selected by two dedicated signature-driven triggers called \emph{CalRatio triggers}~\cite{TRIG-2012-02}, which are designed to identify jets that result from neutral LLPs decaying near the outer radius of the ECal or within the HCal.
The triggers make use of the three main characteristics of the displaced jets: they are narrow jets with a high fraction of their energy deposited in the HCal and typically have no tracks pointing towards the jet.
Two trigger paths are followed in this analysis, defined by two CalRatio triggers that differ only in the level-1 (L1) trigger selection.
The \emph{high-\et trigger} was originally designed for LHC Run 1. The trigger definition was adapted to the Run 2 higher energy and pile-up conditions by, among other modifications, raising the transverse energy (\et) threshold as specified below. This higher threshold has a negative impact on the efficiency for models with $\mPhi\leq200\gev$. To recover efficiency for those models, a new trigger, called the \emph{low-\et trigger}, was designed with a lower threshold.

At L1, the \emph{high-\et trigger} selects narrow jets which each deposit $\et >60\gev$ in a $0.2 \times 0.2~ (\Delta \eta \times \Delta \phi)$ region of the ECal and HCal combined~\cite{ATLAS-CONF-2017-061}.
In September 2016 an upgraded L1 trigger component, the topological trigger, was commissioned in ATLAS. It introduces a new group of triggers that include geometric and kinematic selections on L1 objects.
The \emph{low-\et trigger} makes use of this L1 topological selection by accepting events where the largest energy deposit (and second-largest, if there is one) is required to have $\et > 30\GeV$ deposited in the HCal, with the additional condition that there are no energy deposits in the ECal with $\et>3\GeV$ within a cone of size $\Delta R=0.2$ around the HCal energy deposit. This veto on ECal deposits ensures a high value of $\calrat$ at L1, rejecting a large portion of background events. The trigger rate obtained with this condition is low enough to allow the \et threshold to be kept as low as $30~\GeV$.
This looser \et requirement increases the efficiency for the low-\mPhi signal models (those with $\mPhi\leq200\gev$).

In the high-level trigger (HLT), the selection algorithm for the CalRatio triggers is the same regardless of the L1 selection.
Calorimeter deposits are clustered into jets using the anti-$k_{t}$ algorithm~\cite{Cacciari:2008gp} with radius parameter $R=0.4$.
The standard jet cleaning requirements~\cite{ATLAS-CONF-2015-029} applied in most ATLAS analyses reject jets with high values of $\calrat$, one of the main characteristics of the displaced hadronic jets, and are therefore not included in these triggers.
A dedicated cleaning algorithm for jets created in the HCal (referred to as \emph{CalRatio jet cleaning}) is applied instead, with no requirements on the jet $\calrat$.
At least one of the HLT jets passing the CalRatio jet cleaning is required to satisfy $E_{\mathrm{T}} > 30$~\GeV, $|\eta| < 2.5$ and $\lograt > 1.2$.
Jets satisfying these requirements are used to determine $0.8 \times 0.8$ regions in $\Delta \eta \times \Delta\phi$ centred on the jet axis in which to perform tracking.
Triggering jets are required to have no tracks with $p_{\mathrm{T}} > 2$~\GeV\ within $\Delta R = 0.2$ of the jet axis.
Finally, jets satisfying all of the above criteria are required to pass a BIB removal algorithm that relies on cell timing and position.
Muons from BIB enter the HCal horizontally and may radiate a photon via bremsstrahlung, generating an energy deposit that may be reconstructed as a signal-like jet.
Deposits due to BIB are expected to have a very specific time distribution~\cite{DAPR-2012-01}.
The algorithm identifies events as containing BIB if the triggering jet has at least four HCal-barrel cells at the same $\phi$ and in the same calorimeter layer with timing consistent with that of a BIB deposit.
In both CalRatio triggers, events identified as BIB by the BIB algorithm are saved in the BIB dataset and events with no triggering jets identified as BIB are saved in the main dataset.

The trigger is also active in so-called {\itshape empty bunch crossings}. These are crossings where protons are absent in both beams and isolated from filled bunches by at least five unfilled bunches on either side. Events in empty bunch crossings that have at least one $0.2 \times 0.2~ (\Delta \eta \times \Delta \phi)$ calorimeter energy deposit with $\et>30\GeV$ at L1, and which pass the HLT selection algorithm, are stored in the cosmic-ray dataset.

\begin{figure}[t]
\centering
\includegraphics[width=0.497\linewidth]{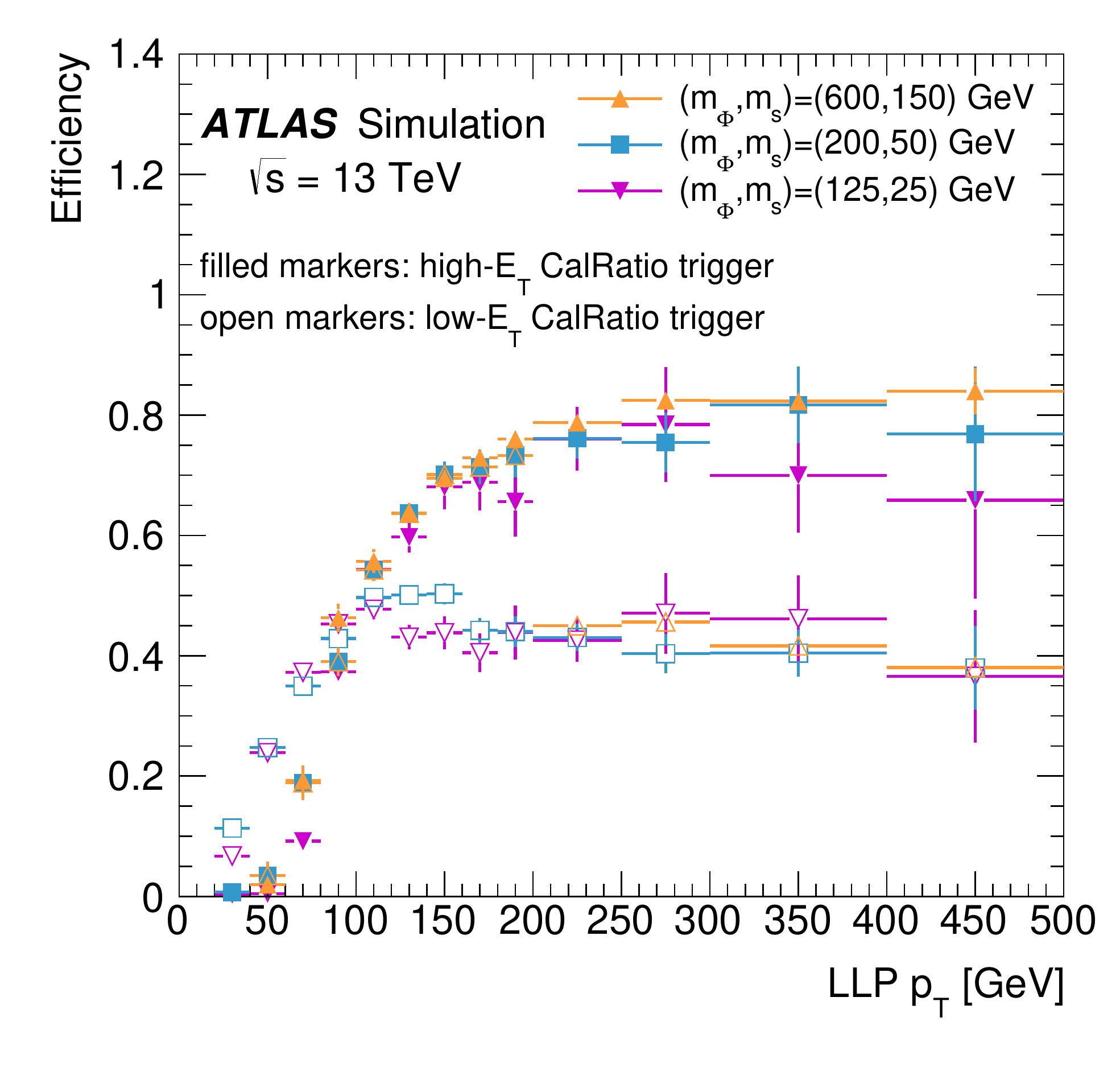}
\includegraphics[width=0.497\linewidth]{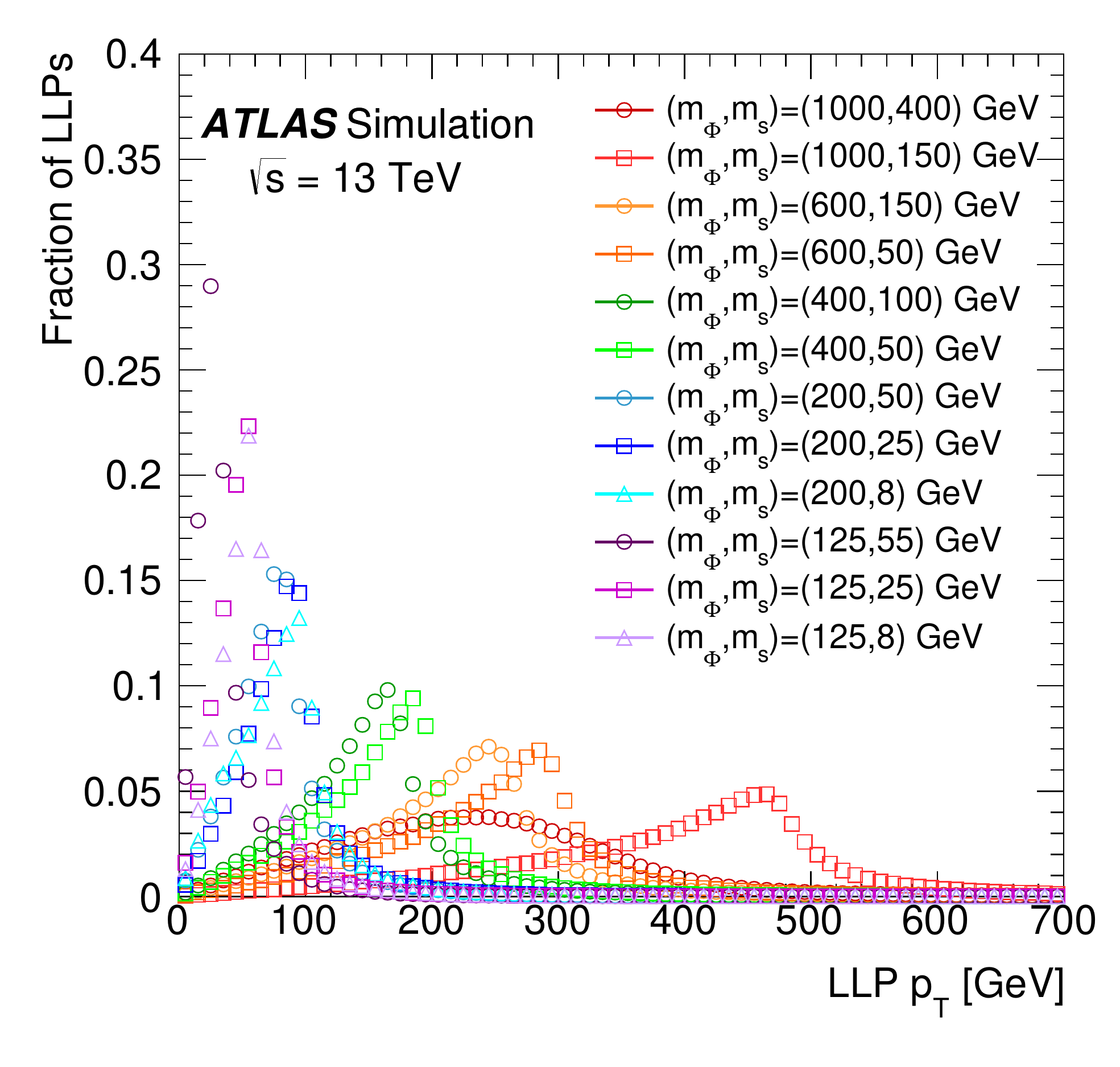}
\caption{Trigger efficiency of simulated signal events as a function of the LLP \pt\ (left) and the \pt\ distribution of LLPs (right) for a selection of signal samples.
}
\label{fig:trigEffpt}
\end{figure}

The trigger efficiency for simulated signal events is defined as the fraction of jets spatially matched to one of the generated LLPs (hereafter called \emph{truth LLPs}) that fire the trigger.
The trigger efficiency as a function of triggering LLP particle-level \pt is shown in Figure~\ref{fig:trigEffpt} (left) for two signal samples.
Only LLPs decaying in the HCal are considered in this plot.
The high-\et CalRatio trigger, which is seeded by the high-\et L1 trigger, starts to be efficient for LLPs with $\pt > 100\GeV$ and reaches its plateau at 150--200~\GeV.
The low-\et CalRatio trigger (seeded by the low-\et L1 trigger) recovers efficiency for a large portion of the LLPs with $\pt < 100\GeV$.
The main source of efficiency loss in these triggers comes from the track isolation, followed by the combination of requirements on jet \et and \calrat.
Figure~\ref{fig:trigEffpt} (right) shows the LLP \pt distribution for all the signal samples considered in the analysis.
The combination of these figures shows how the high-\et CalRatio trigger gives a higher efficiency for models with $\mPhi > 200\GeV$, where the LLP $\pt$ distributions peak between 150 and $500\gev$.
For signal models with $\mPhi$ up to $200\GeV$, the LLP $\pt$ distributions peak between $30$ and $100\gev$ and hence the low-\et CalRatio trigger performs better.
Thus, low-\mPhi models are searched for using the low-\et dataset: despite the reduced integrated luminosity, a higher sensitivity is obtained than if the high-\et dataset had been used. Conversely, models with $\mPhi > 200\gev$ are studied using the high-\et dataset.

\begin{figure}[ht]
\centering
\includegraphics[width=0.497\linewidth]{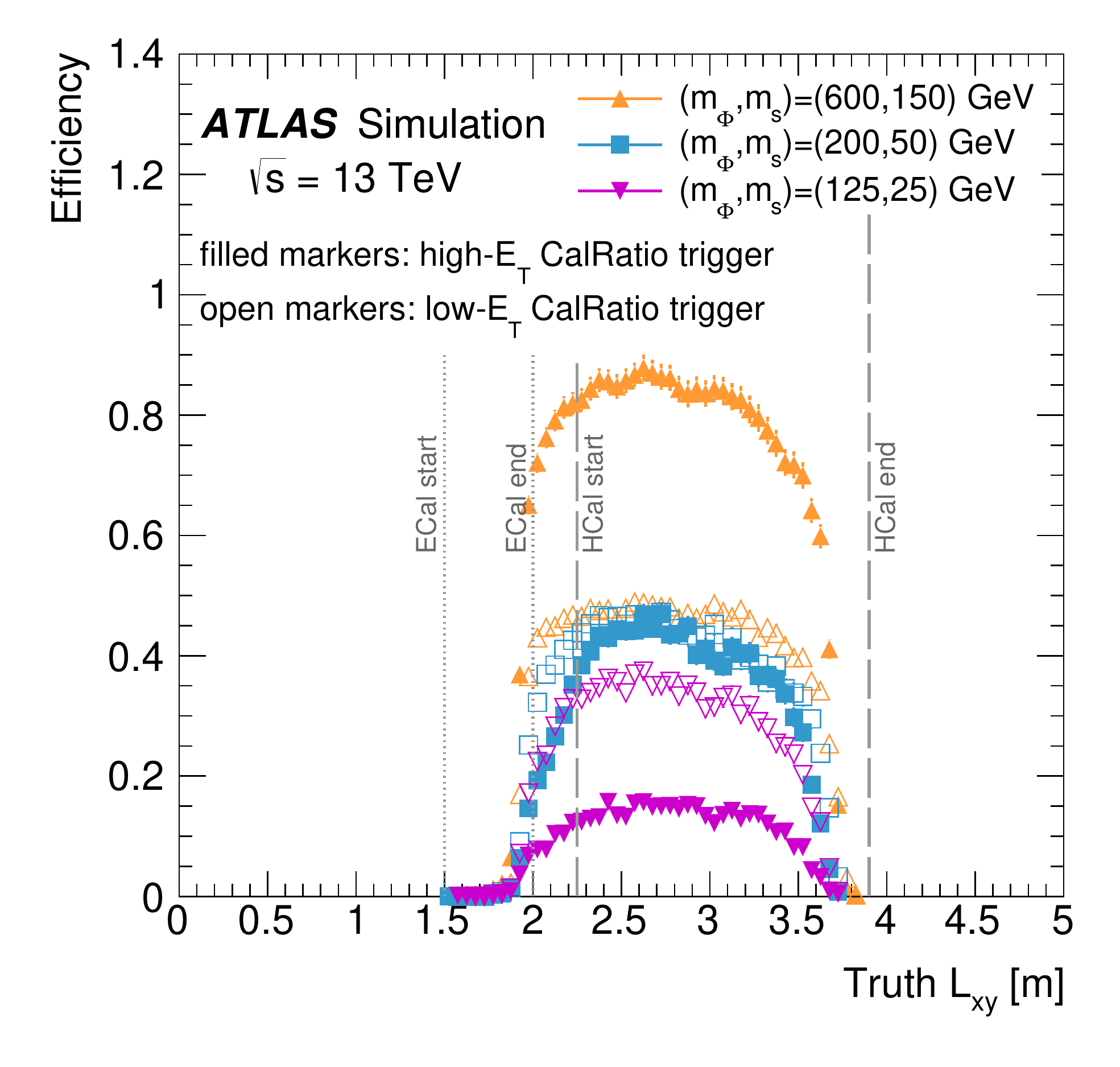}
\includegraphics[width=0.497\linewidth]{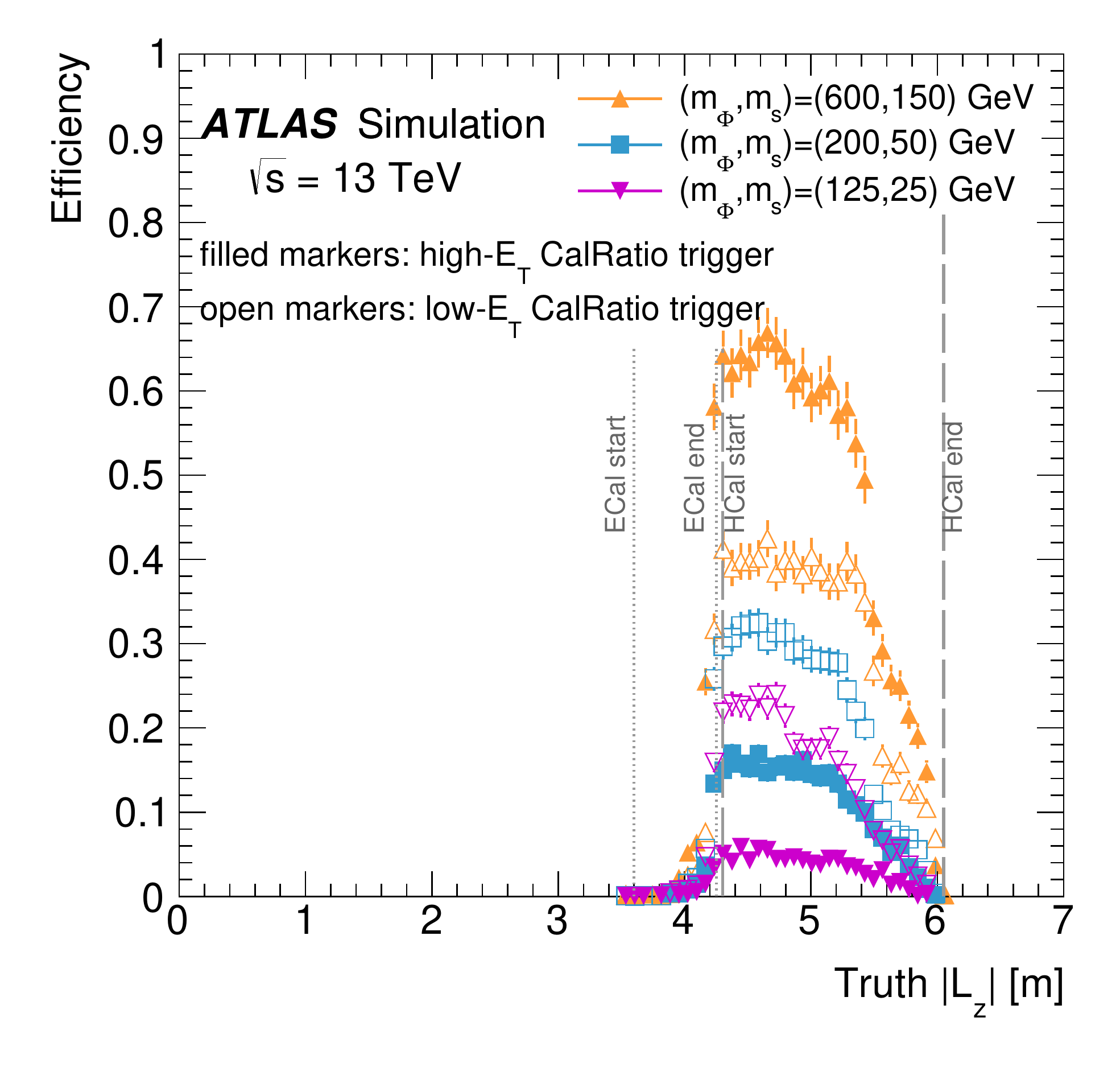}
\caption{Trigger efficiency of simulated signal events as a function of the LLP decay position in the $x$-$y$ plane for LLPs decaying in the barrel (left, $|\eta|<1.4$) and in the $z$ direction for LLPs decaying in the HCal endcaps (right, $1.4 \leq |\eta| < 2.5$) for three signal samples. The open (filled) markers represent the efficiency for events passing the low-\et (high-\et) CalRatio trigger.
}
\label{fig:trigEffLxy}
\end{figure}

The trigger efficiency also depends strongly on the LLP decay position, as shown for three samples of simulated signal events in Figure~\ref{fig:trigEffLxy}.
The efficiency as a function of LLP decay length in the $x$--$y$ plane is shown for LLPs decaying in the barrel ($|\eta|<1.4$); the efficiency as a function of the decay position in the $z$-direction is shown for LLPs decaying in the HCal endcaps ($1.4 \leq|\eta| < 2.5$).
The selection is most efficient in the HCal for both triggers.

Events used in the analysis are required to pass the trigger requirements and contain a primary vertex (PV) with at least two tracks with $\pt > 400$~MeV. 
Tracks used in the jet and event selection hereafter are required to pass the \emph{track selection}: they must originate from the PV and have $\pT>2\gev$.

The jets used in this analysis are selected by applying the following quality selections: $\pt > 40~\GeV$, $|\eta| < 2.5$, pass CalRatio jet cleaning. These jets are referred to as \emph{clean}.
To select events with trackless jets, an additional event-level variable, $\sumMinDR{\mathrm{(jet, tracks)}}$, is used.
The quantity $\Delta {R}_\mathrm{min}{\mathrm{(jet, tracks)}}$ is defined as the angular distance between the jet axis and the closest track with $\pt>2\GeV$,
and $\sumMinDR{\mathrm{(jet, tracks)}}$ is calculated by summing this distance over all the clean jets with $\pt>50\GeV$.
Events with no displaced decays have a very small value of this variable. Every displaced jet contributing to the sum causes a considerable increase in the value, making this variable a good discriminator between signal and multijet background.
For an event to pass the analysis \emph{preselection}, it is required to have passed the trigger, to contain at least two clean jets and to have $\sumMinDR{\mathrm{(jet, tracks)}}>0.5$.
After preselection, $\sumMinDR{\mathrm{(jet, tracks)}}$ still has good discrimination power and it is used in the data-driven background estimation described in Section~\ref{sec:background}.

\subsection{Displaced jet identification}
\label{sec:DJID}

Each clean jet is evaluated by a multilayer perceptron (MLP) (implemented in the Toolkit for Multivariate Data Analysis~\cite{TMVA}) to predict the radial and longitudinal decay positions (\lxy and \lz) of the particle that produced the jet, using the jet's fraction of energy deposited in each of the ECal and HCal layers as input variables. The MLP was trained on simulated signal samples with $\mPhi$ in the range $[200,1000]\GeV$, using only jets matched to a truth LLP.
No requirements at event level (trigger and preselection) were applied in order to have as large a data sample as possible.
In addition, avoiding the preselection allows the MLP to identify the decay position of prompt jets, which is useful when applied to SM jets.
The MLP training procedure took as input the truth-level \lxy and \lz decay positions of the LLP as well as the fraction of the jet energy in each calorimeter layer, and finally the jet's direction in $\eta$.

\begin{figure}[t]
\centering
\includegraphics[width=0.497\linewidth]{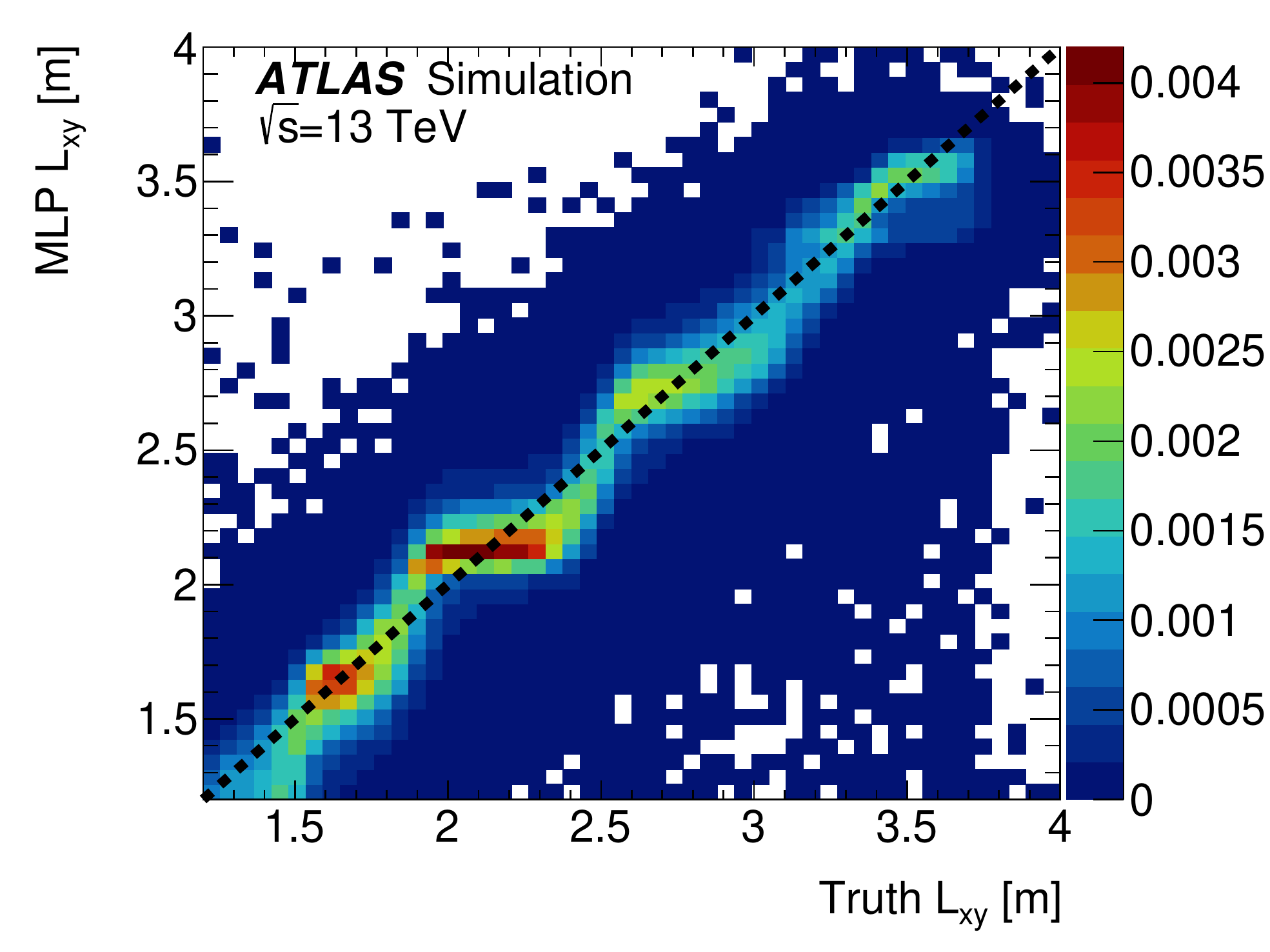}
\includegraphics[width=0.497\linewidth]{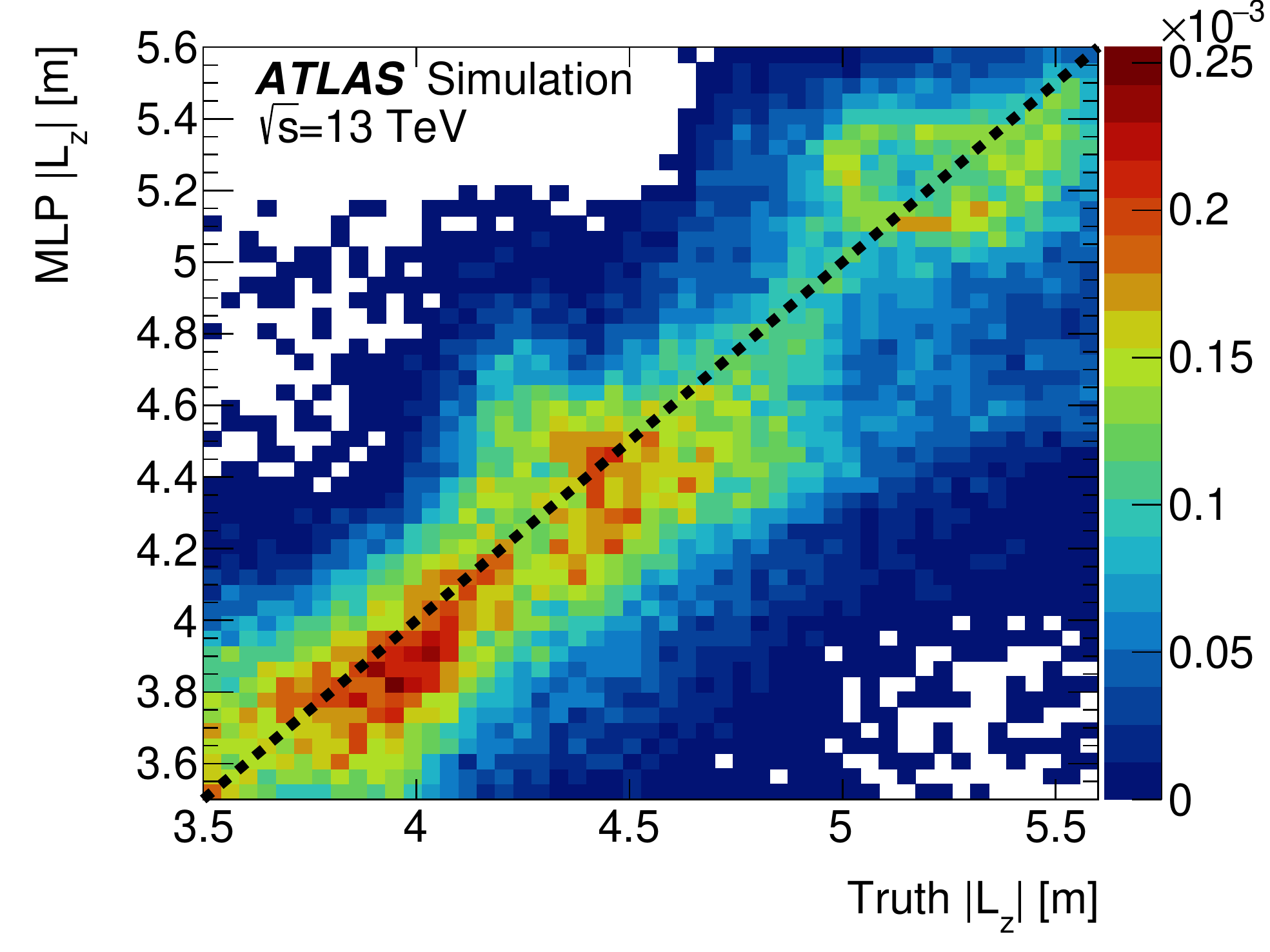}
\caption{Probability density of predicted MLP radial (\lxy, left) and longitudinal (\lz, right) LLP decay positions as a function of the truth LLP decay positions, for reconstructed jets matched to the LLP. Dotted lines show where the MLP value equals the truth value.
}
\label{fig:jetMLP}
\end{figure}

The left-hand plot of Figure~\ref{fig:jetMLP} compares \lxy of a truth LLP against the MLP prediction. It shows clearly the different calorimeter layers, since decays in the same layer lead to constant MLP radial decay position prediction even as the truth decay position changes. However, the overall prediction in \lxy aligns closely with the truth decay position. The right plot shows the longitudinal decay position, \lz. It shows a clear correlation between prediction and truth for the whole range of the forward calorimeter with less obvious layering, since the LLP direction of travel in the endcaps is more oblique with respect to the calorimeter layers than in the barrel.
The radial and longitudinal decay positions predicted by the MLP are useful discriminators between signal jets from LLP decays in the calorimeters and prompt jets from SM backgrounds.

The per-jet BDT is used to separate jets into three classes: signal-like jets, SM multijet-like jets and BIB-like jets.
With that purpose, it is trained using three samples.
The signal sample contains jets from signal events for a range of models with $\mPhi$ in the range $125$ to $1000 \GeV$, where only jets matched to LLPs decaying outside the ID (with $\lxy > 1250$~mm if they decay in the barrel or $\lz > 3500$~mm if they decay in the endcaps) are considered.
The SM multijet training sample consists of jets from the simulated multijet events described in Section~\ref{sec:sig_bkgMC}.
Finally, the BIB sample is made of jets from the BIB dataset, where only the triggering jet in each event is used. The triggering jet is identified as BIB by the trigger BIB algorithm: the event contains a line of at least four HCal-barrel cells in the same $\phi$ as the triggering jet, consistent with BIB timing.
Hence, the triggering jet corresponds to a BIB jet in most cases, which is confirmed by the $\phi$ and $z$ vs. time plots showing the typical shapes of BIB.
Using only the triggering jet reduces the risk of contamination from multijet events.
In all cases, only clean jets are considered.

The per-jet BDT inputs are the MLP \lxy and \lz predictions, track variables, and jet properties.
The track variables include the sum of \pt of all tracks passing track selection within $\Delta R = 0.2$ of the jet axis, and the maximum \pt of such tracks.
The jet properties are: the radius, shower centroid, energy density and fraction of energy in first HCal layer of the cluster with the highest \pt; the longitudinal and transverse distance from this cluster to the jet shower center; jet \pt ; and the compatibility of the jet timing with the expected timing of a BIB deposit.

The jet $\pt$ spectrum is very different in each of the three training samples, and therefore jets in each sample are weighted such that the jet \pt distribution is flat. The weighting is done independently in each training sample. Since the jet $\pt$ is correlated with a number of BDT input variables, the jet $\pt$ is also included as a variable in the BDT.

\begin{figure}[ht]
\centering
\includegraphics[width=0.48\linewidth]{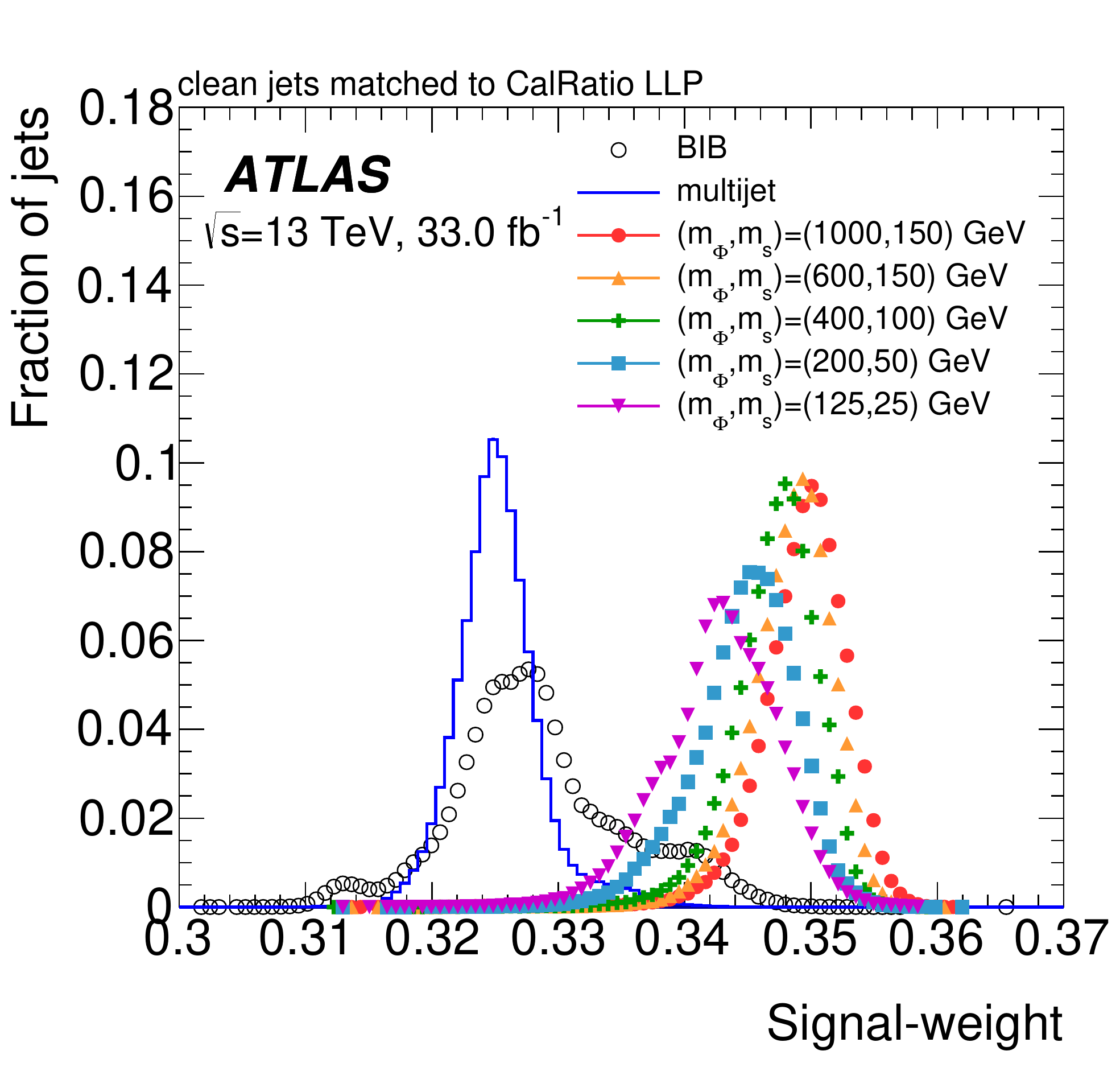}
\includegraphics[width=0.48\linewidth]{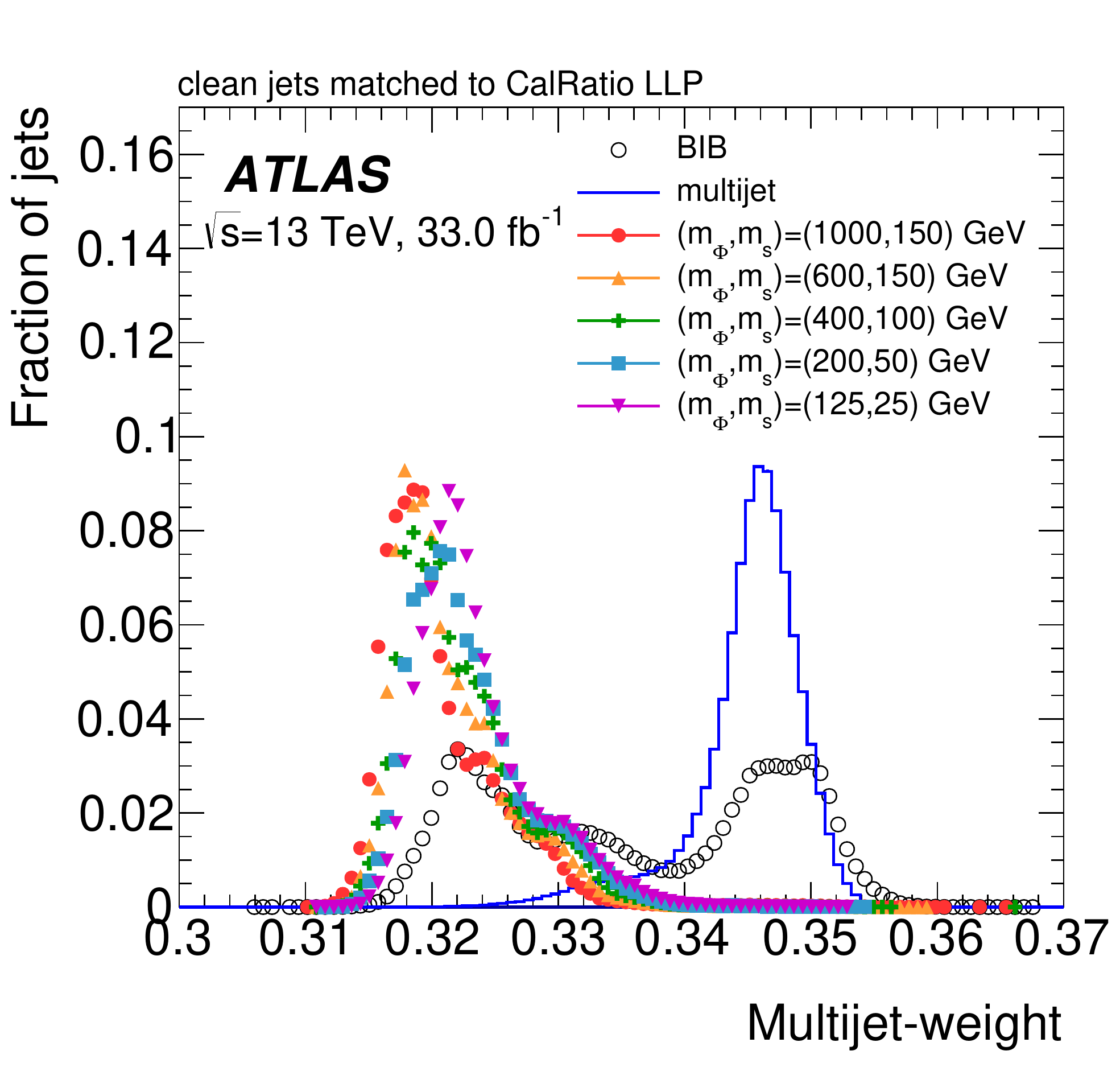}\\
\includegraphics[width=0.48\linewidth]{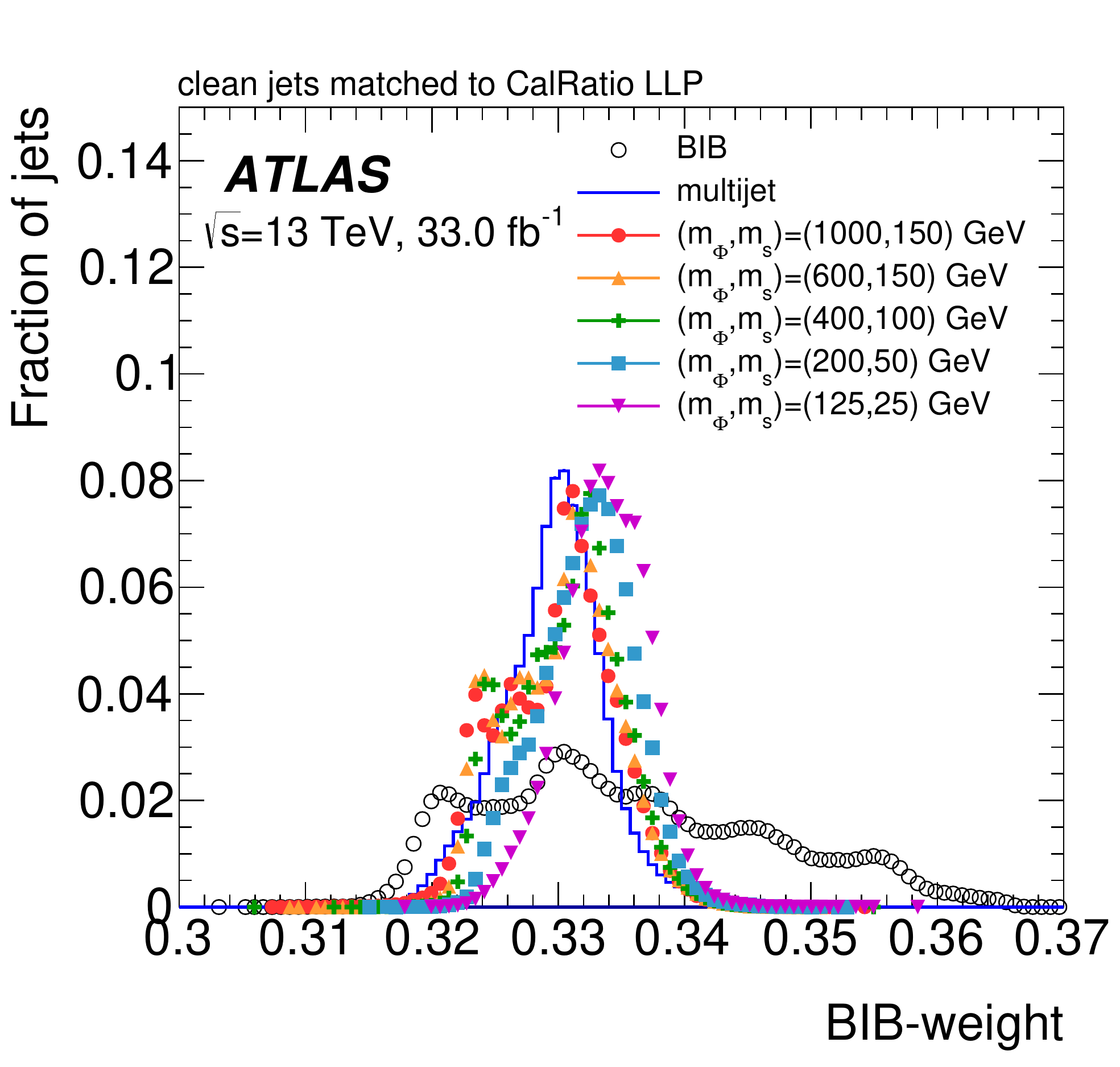}
\caption{The distributions of the per-jet BDT weights for a multijet sample, a BIB sample and five signal samples. For the signal samples, the weights for clean jets matched to an LLP decaying in the calorimeter are shown. The multijet and BIB distributions include weights for all clean jets in the event.
}
\label{fig:perjetBDTweights}
\end{figure}

The output of the per-jet BDT is a set of three weights that sum to unity: signal-weight, BIB-weight and multijet-weight, shown in Figure~\ref{fig:perjetBDTweights}. The signal-weight distribution provides a clear separation between signal jets and both types of background jets.
The BIB-weight distributions for signal and multijet jets peak at intermediate values.
Jets from the BIB sample with low BIB-weight scores ($<0.34$) display SM multijet-like qualities and are likely to result from SM jet contamination in the BIB sample. Jets with higher BIB-weight values display the expected timing behaviour of particles originating from BIB. The per-jet BDT is able to separate these with some precision, assigning values between 0.34 and 0.35 to BIB particles crossing the detector through the innermost layer of the HCal and higher values ($>0.35$) to BIB in outer HCal layers.

The per-jet BDT has better signal-to-background discrimination for high-\mPhi models than for low-\mPhi models.
The main reason for this lies in the \pt distribution (see Figure~\ref{fig:trigEffpt}). Both the BIB and pile-up jets have relatively soft \pt, and even though these backgrounds are mitigated by the jet-cleaning requirements, their remaining contributions are harder to distinguish at low \pt.
The presence of pile-up jets has two effects: on the one hand, they can leave energy deposits in the ECal, changing the fraction of energy per calorimeter layer and worsening the signal-to-background discrimination.
On the other hand, pile-up jets' tracks do not point back to the PV in many cases and hence are not considered for track isolation. These jets can be reconstructed as nearly trackless, making them more similar to signal.

\subsection{Event selection}
\label{sec:evSelection}

A per-event BDT is defined with the main objective of discriminating BIB events from signal events.
A combination of signal samples is used as signal in the training while the BIB dataset events are used as background.

The two jets with the highest per-jet signal-weight in the event (\emph{CalRatio jet candidates}) and the two jets with the highest per-jet BIB-weight in the event (\emph{BIB jet candidates}) are selected and their per-jet weights are used as input variables to the per-event BDT.
Other event-level variables such as \mhtoht , where $H_{\mathrm T}$ is the scalar sum of jet transverse momenta and $\mht$ is the magnitude of the vectorial sum of transverse momenta of these jets, and the distance $\Delta R$ between the two CalRatio jet candidates are used in the training.

\begin{figure}[ht]
\centering
\includegraphics[width=0.497\linewidth]{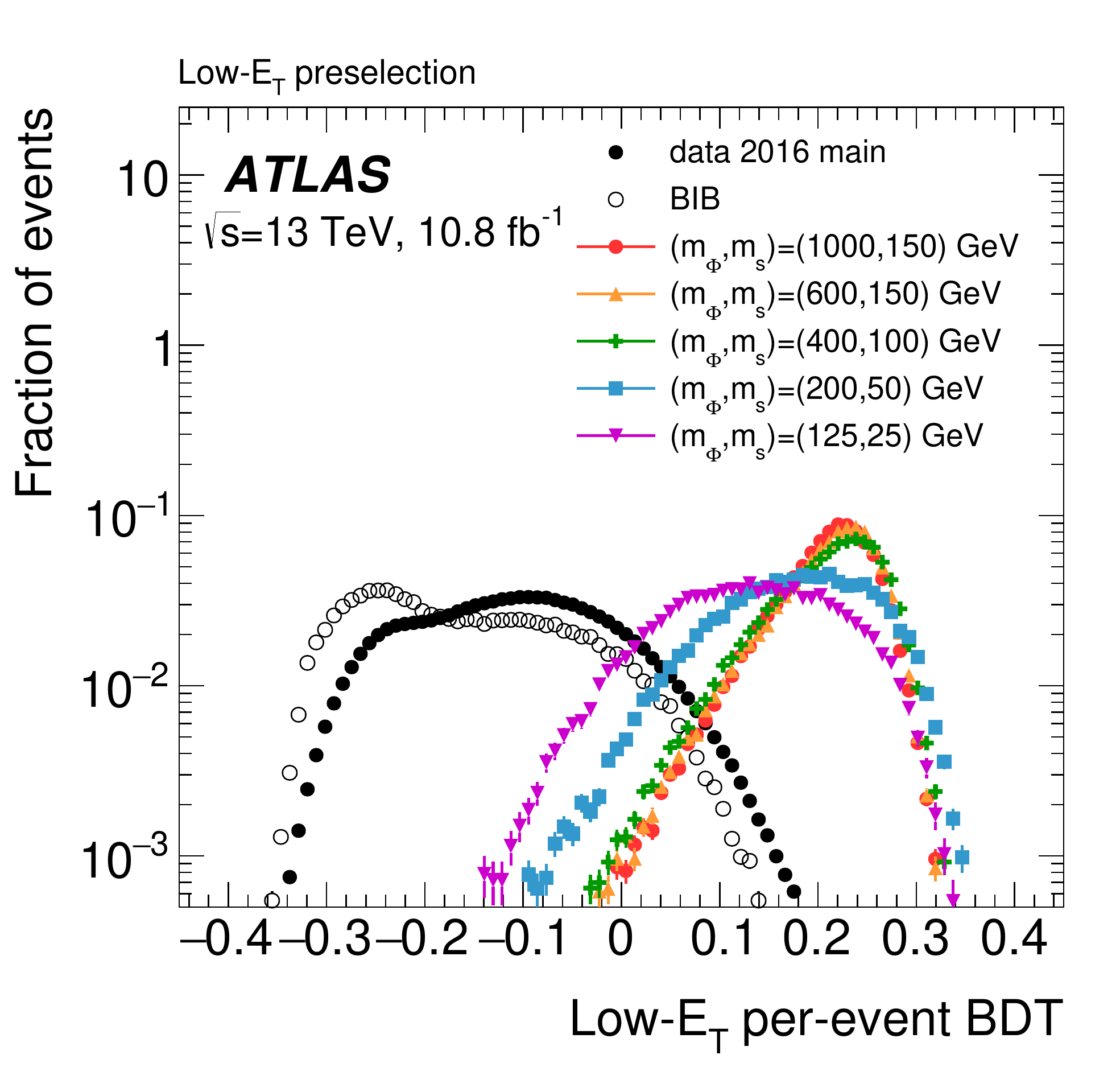}
\includegraphics[width=0.497\linewidth]{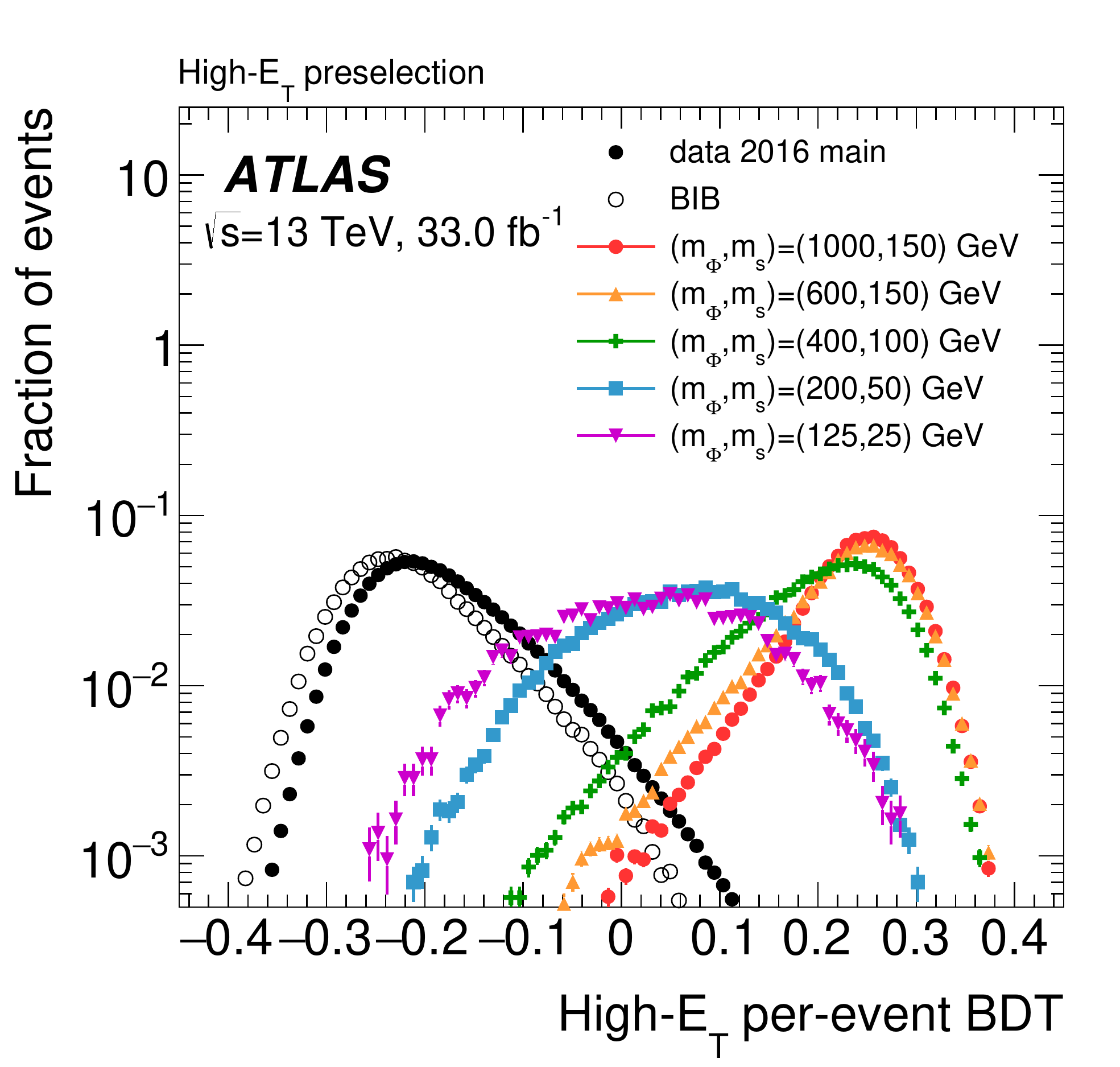}
\caption{Distribution of the low-\et per-event BDT (left) and high-\et per-event BDT (right) on main data, BIB data and five signal samples after preselection.
}
\label{fig:eventBDT}
\end{figure}

As mentioned in the previous subsection, signal jets with low \pt are harder to discriminate from background. For this reason, and to obtain an optimal signal-to-background discrimination at all \pt, two versions of the per-event BDT are trained: one for the analysis of the high-\et dataset, and another for the low-\et dataset.
They only differ in the signal samples used for training and in the triggers required to select events.
The high-\et per-event BDT training uses a combination of low-, intermediate- and high-mass signal samples in events passing the high-\et CalRatio trigger.
The low-\et per-event BDT training uses a combination of low-\mPhi signal samples and only events passing the low-\et CalRatio trigger.
Figure~\ref{fig:eventBDT} shows the distribution of the per-event BDTs from five signal samples, as well as from the main data and BIB data.
The BIB training sample contains SM multijet jets in addition to the BIB jet that caused them to be selected by the trigger.
Consequently, even if no multijet sample is used in the training, the per-event BDT is able to discriminate signal from BIB as well as from multijet background. This can be seen in Figure~\ref{fig:eventBDT} by comparing the BDT results in the main data and the BIB datasets, especially in the low-\et per-event BDT output. Using time and z-coordinate measurements, it has been checked that events with low per-event BDT values ($<-0.2$) have the typical characteristics of BIB, while events with intermediate values (between $-0.2$ and $0.2$) are multijet-like.

The simulated distributions of the variables used as BDT inputs (for both the per-jet and per-event BDTs) are compared with data, and good agreement is generally observed. The small remaining discrepancies are propagated into an uncertainty in the modelling of BDT input variables, which is described in Section~\ref{sec:systematics}.

Two selections are defined, referred to as the \emph{high-\et selection} and the \emph{low-\et selection}, which are optimised to give maximum sensitivity for high-\mPhi models and low-\mPhi models, respectively.

\emph{Event cleaning} selections are applied to remove as much BIB background as possible: trigger matching (at least one of the CalRatio jet candidates has to be matched to the jet that fired the trigger), and a timing window of $-3 < t < 15$~\ns for the CalRatio jet candidates and for the BIB jet candidates. Furthermore, the per-event BDT output is required to satisfy \evbdtLITau$>0.1$ and \evbdtLITopo$>0.1$ in the high-\et and low-\et selections, respectively. These requirements ensure that the only source of background contributing to the final selection is multijet events.

The final selection is optimised to maximise the signal-to-background ratio in each search region. Variables with good signal-to-background discrimination at event level are used, such as \mhtoht and $\sumLogR$, where \FCRjet and \SCRjet refer to the CalRatio jet candidates.
The quantity \mhtoht has a value close to 1 for BIB events, but it has a softer distribution for signal.
This variable replaces the $\met <  30~\GeV$ requirement applied in the 8 TeV analysis~\cite{EXOT-2012-28} (where $\met$ is the magnitude of the negative vector transverse momentum sum of the reconstructed and calibrated physics objects), which was very useful for reducing the multijet background with only a small effect on the efficiency of low-\mPhi models. However, it significantly lowered the efficiency for the high-\mPhi models due to larger portions of the high-\pt jets escaping the calorimeters (\emph{punch-through}), generating fake \met.
The elimination of this requirement improves the sensitivity of the analysis to the high-\mPhi models by a large factor, while the improvement is less noticeable for low-\mPhi.
The following additional requirements are applied for the high-\et selection: $\sumLogR > 1$, $\FCRjetpt> 160\gev$, $\SCRjetpt> 100\gev$, and $\mhtoht <0.6$. The low-\et selection requires $\sumLogR > 2.5$, $\FCRjetpt>80\gev$, and $\SCRjetpt>60\gev$.

\section{Background estimation}
\label{sec:background}

The data-driven ABCD method is used to estimate the contribution from the dominant background (SM multijet events) to the final selection.
The standard ABCD method relies on the assumption that the distribution of background events can be factorised in the plane of two relatively uncorrelated variables.
In this plane, the method uses three control regions (B, C and D) to estimate the contribution of background events in the search region (A).
If all the signal events are concentrated in region A, the number of background events in region A can be predicted from the population of the other three regions using $N_\mathrm{A}=(N_\mathrm{B}\cdot N_\mathrm{C})/N_\mathrm{D}$, where $N_{X}$ is the number of background events in region $X$.
In reality, some signal events may lie outside of region A. A modified ABCD method is used to account for non-zero signal contamination in regions B, C and D. The modified ABCD method involves fitting to background and signal models simultaneously. The background component of the yields in regions A, B, C and D are constrained to obey the standard ABCD relation, within the bounds of the ABCD method uncertainty (described below).
In the modified ABCD method, the signal strength is also included as a parameter in the fit, which may uniformly scale the signal yield in each region.
The good performance of the method is only ensured in the presence of a single source of background. In this case the background must be confirmed to be dominated by SM multijet events.
Two checks are performed to ensure that the contribution of background events from non-collision background after the selection is negligible.
The fraction of events satisfying each stage of the selection for the main data, BIB background, cosmic-ray background and benchmark signal samples is shown in Table~\ref{tab:Selection_cutFlow} for the high-\et and low-\et selections.

First, the number of BIB events passing each stage of the analysis selections is checked.
For both the high-\et and low-\et selections, the number of BIB events satisfying all selection criteria is well within the uncertainty in the number of events passing all selections in the main dataset. Furthermore, the events from the BIB dataset that pass the selection were checked, and found to display properties of multijet events.
In particular, their $\phi$ and $z$ vs time distributions do not show the typical shape of BIB.
The events from the main dataset that pass the event cleaning were also checked, and were found not to display the properties of BIB.

The second check is to ensure that almost all the cosmic-ray background is removed, using the cosmic-ray dataset. The estimated number of events passing each stage of the selection is listed in Table~\ref{tab:Selection_cutFlow} for the high-\et (low-\et) selection.
In both cases the number is also within the statistical uncertainty for the number of events entering the selection in the main dataset.

\begin{figure}[ht]
\centering
\includegraphics[width=0.329\linewidth]{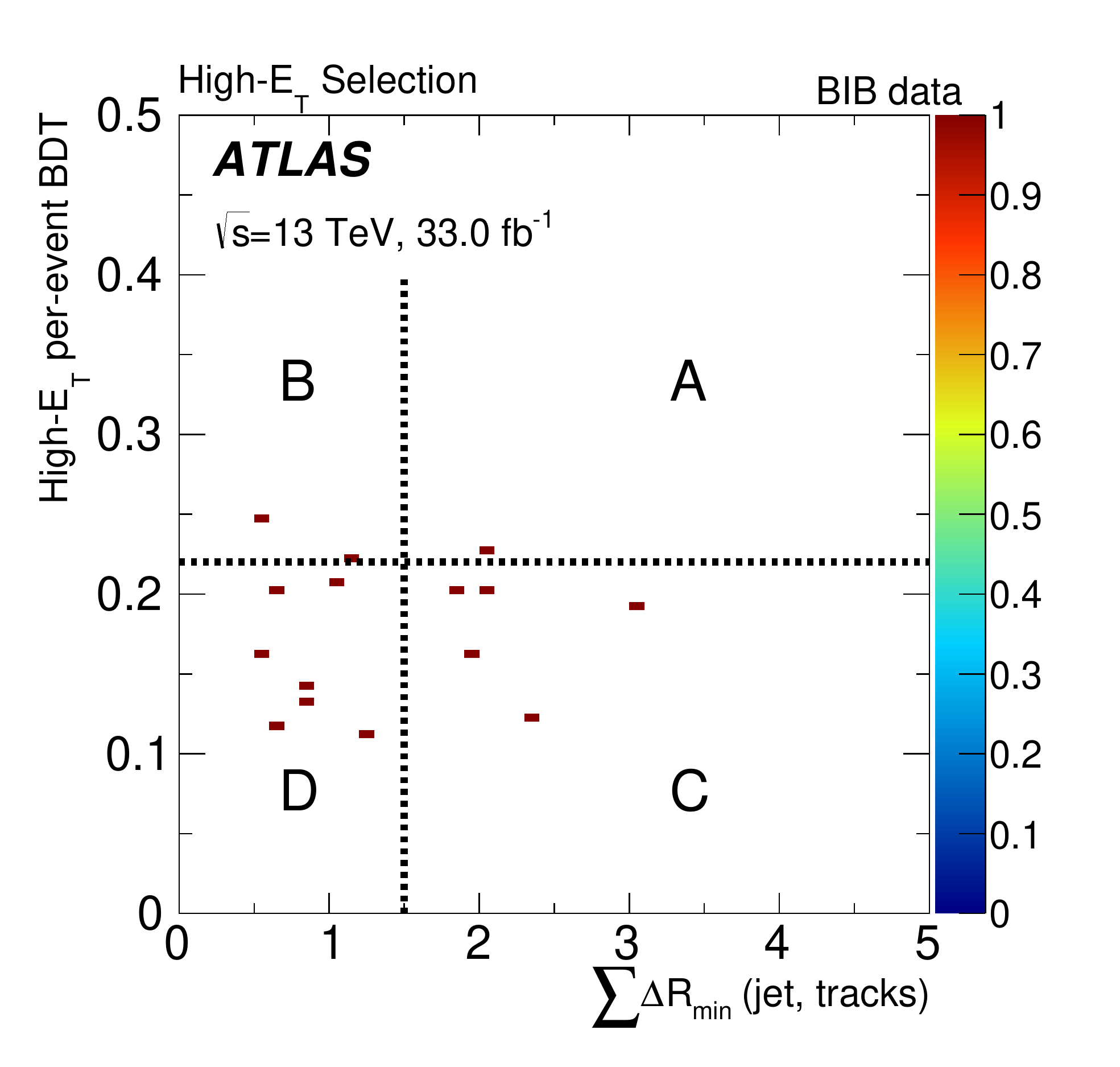}
\includegraphics[width=0.329\linewidth]{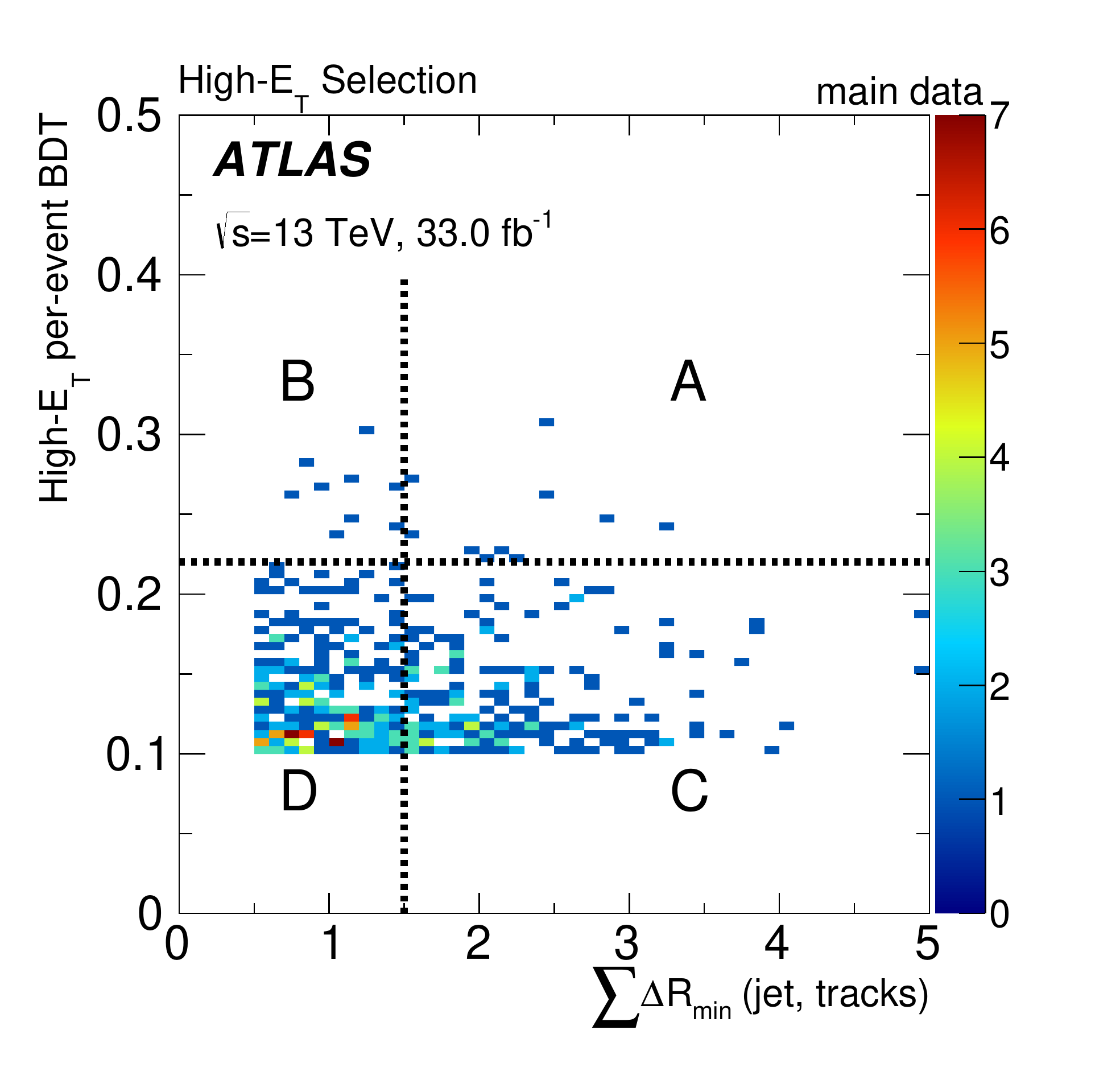}
\includegraphics[width=0.329\linewidth]{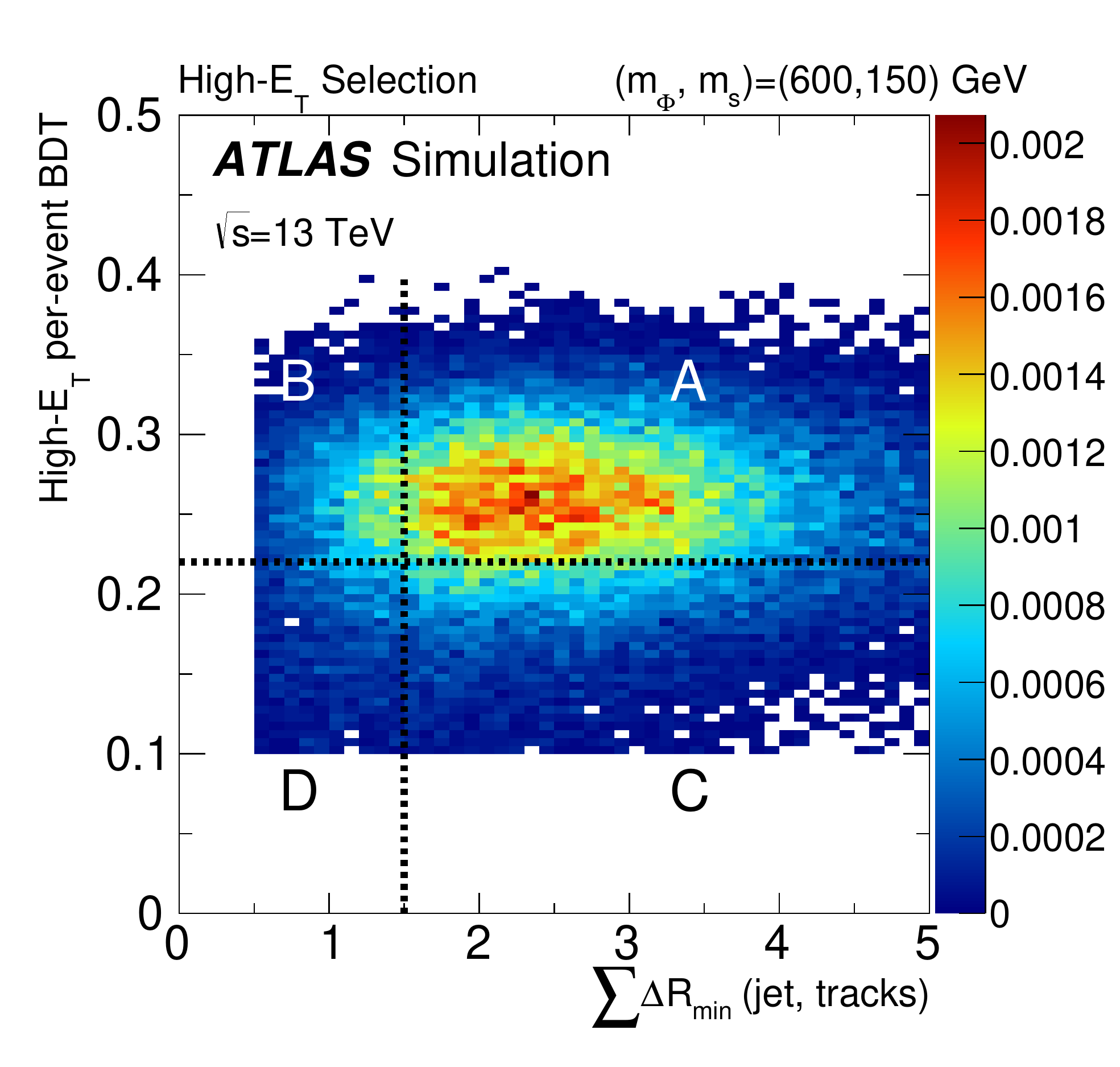}

\includegraphics[width=0.329\linewidth]{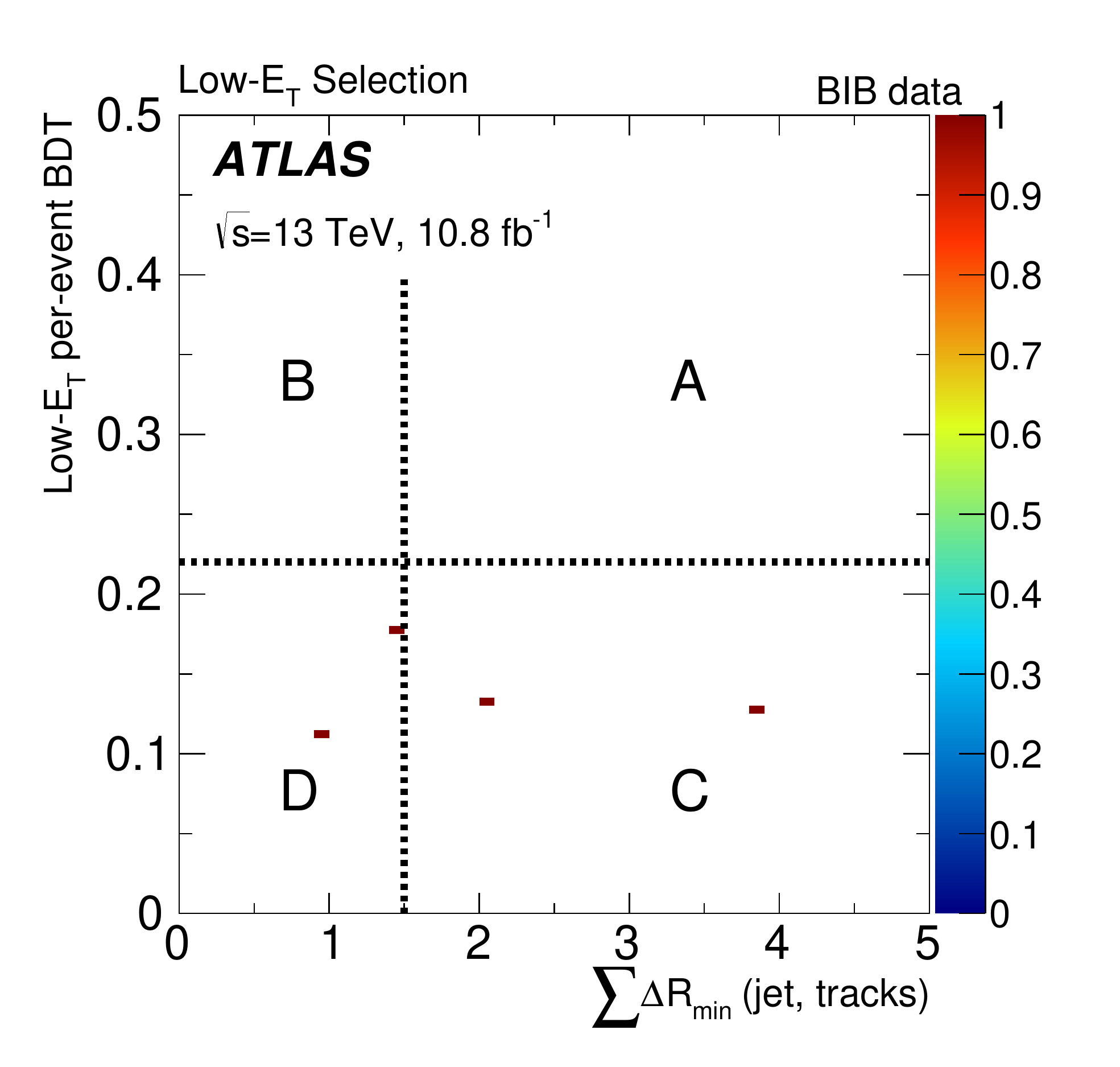}
\includegraphics[width=0.329\linewidth]{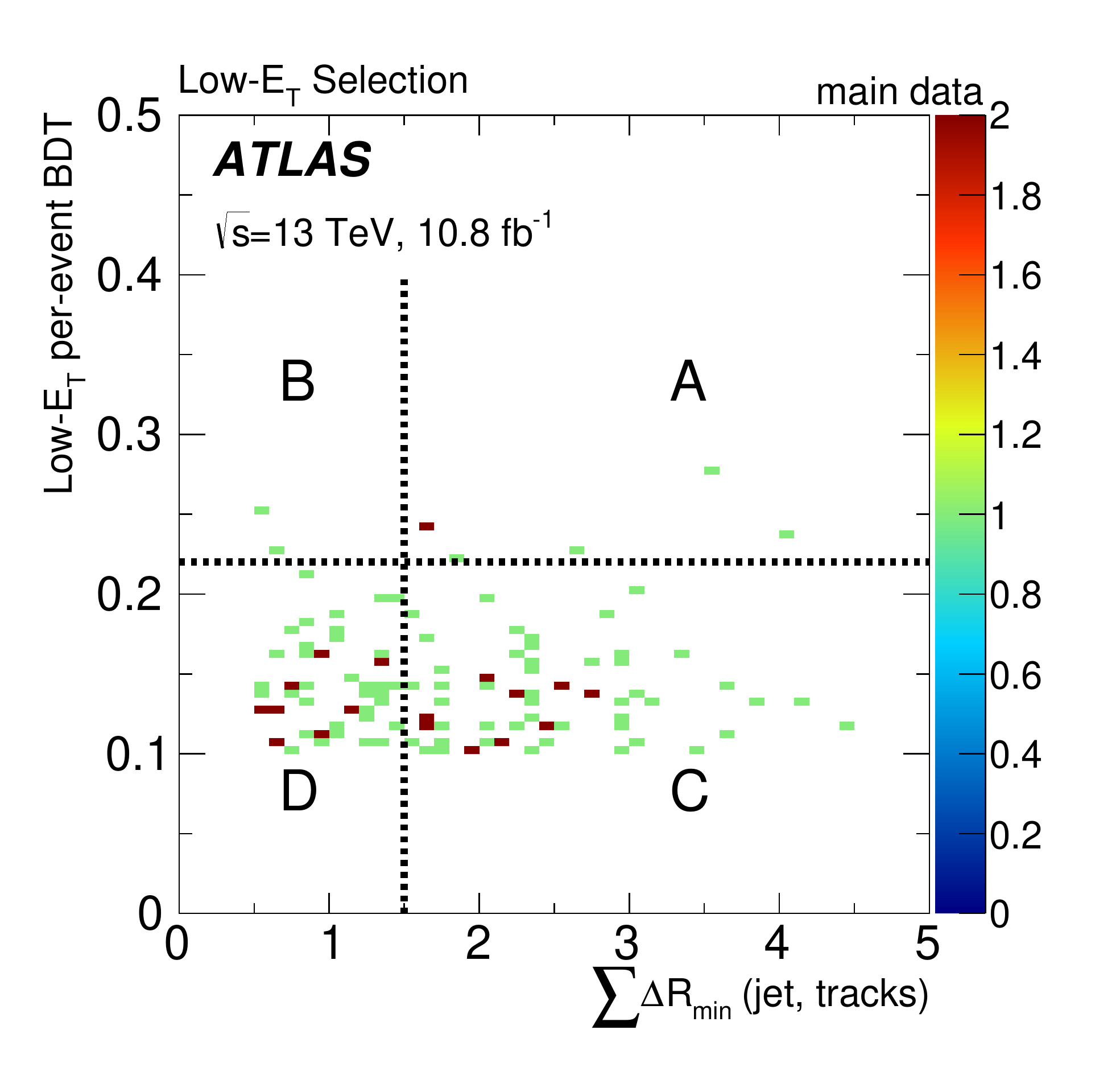}
\includegraphics[width=0.329\linewidth]{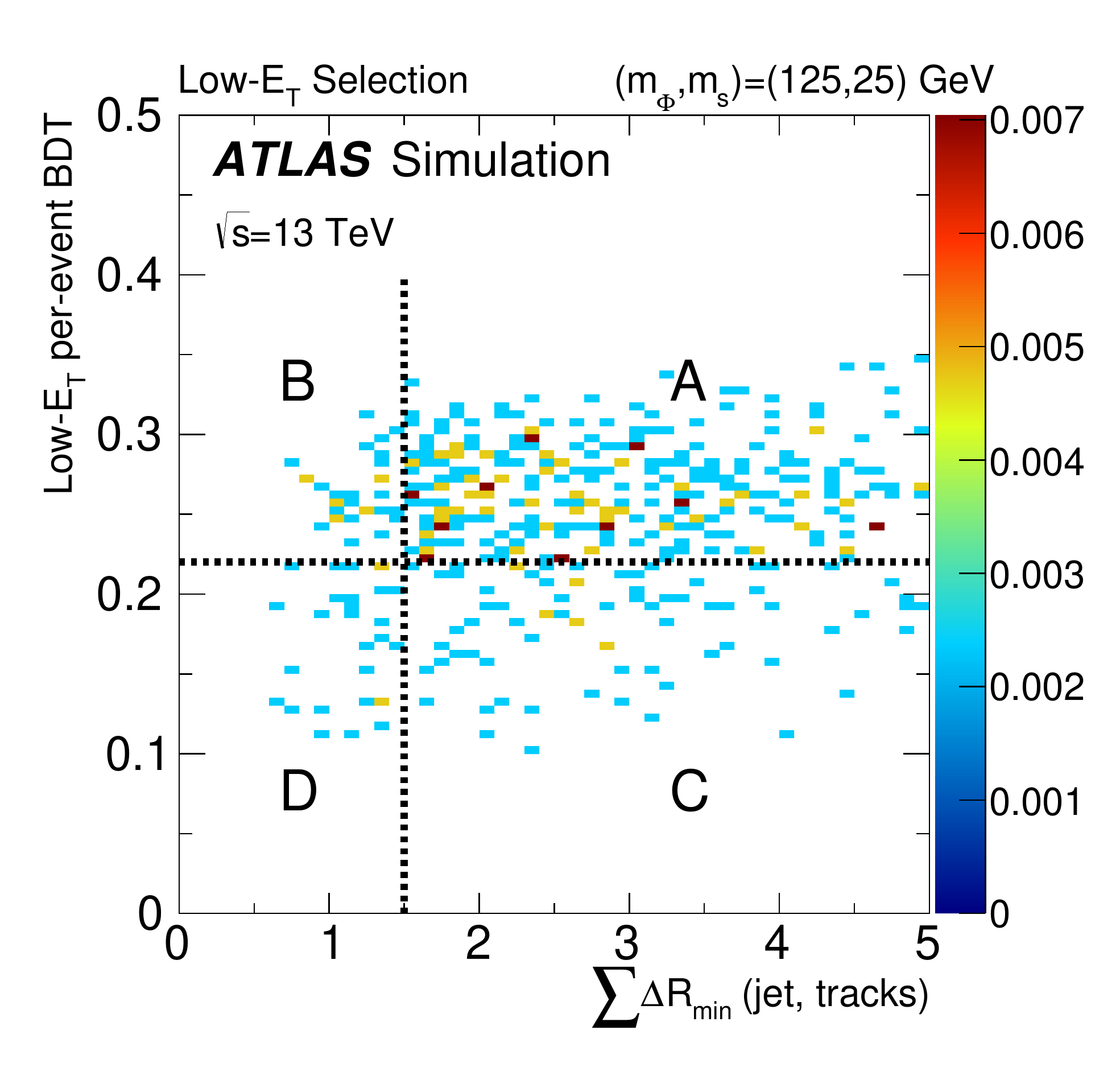}
\caption{The distributions of $\sumMinDR$(jet, tracks) versus \evbdtLITau (top row) and \evbdtLITopo (bottom row) for BIB events (left), main data (centre) and a signal sample (right) after event selection. The signal sample with $\mPhi = 600 \GeV$ and $\mS = 150 \GeV$ is shown for the high-\et selection, while the $\mPhi = 125 \GeV$ and $\mS = 25 \GeV$ sample is shown for the low-\et selection.
Signal plots are shown as a probability density. The black dashed lines indicate the boundaries defining regions A, B, C and D in the plane after event selection.
}
\label{fig:ABCDplane_SR}
\end{figure}

The two variables chosen to form the ABCD plane are $\sumMinDR$(jet, tracks) and \evbdtLITau or \evbdtLITopo, depending on the selection.
The variables are uncorrelated (correlation $< 4\%$ in main data after the event cleaning) and have good separation between signal and multijet background, as shown in Figure~\ref{fig:ABCDplane_SR}.
An optimization procedure is applyied to define the most efficient selection of regions A, B, C and D. Different boudaries are tested to maximise the ratio $\mathrm S\sqrt(B)$ where S is the number of signal events in region A and B is taken as the background estimation given by the ABCD method for each of the studied selections. Only selections with low signal contamination in regions B, C and D are considered.
Following this procedure, region A is defined by $\sumMinDR \geq1.5$ and per-event BDT $ \geq 0.22$ for both the high-\et and low-\et selections.
Regions B, C, and D are defined by reversing one or both of the requirements: ($\sumMinDR <1.5$ and per-event BDT $ \geq 0.22$), ($\sumMinDR \geq1.5$ and per-event BDT $ < 0.22$) and ($\sumMinDR <1.5$ and per-event BDT $ < 0.22$) respectively.
Figure~\ref{fig:ABCDplane_SR} shows the distribution of events in the ABCD plane for the BIB dataset, the main dataset and one representative signal sample, after the final selection is applied. Signal and background events populate different regions in the plane. As a reference, the boundaries defining regions A, B, C and D are indicated in the same figure by black dashed lines.

\begin{landscape}
\begin{table}[h]
\caption{Sequential impact of each requirement on the number of events passing the selection for the high-\et (top) and low-\et (bottom) analyses. The quoted number for BIB corresponds to the number of events in the BIB dataset passing the selection. The number of events for the cosmic rays is estimated from the cosmic-ray dataset by applying corrections for the proportion of empty crossings relative to filled crossings, and a factor to account for the change in analysis efficiency in a zero-pile-up collision. Signal yields are quoted as a percentage of the total number of generated events.}
\resizebox{1.4\textwidth}{!}{
\begin{tabular}{llrrrrrrrr}
\hline
&High-\et selection:  &  Main data &   BIB & Cosmic rays &  Signal $(\mPhi,\mS)$                   &     Signal $(\mPhi,\mS)$     &    Signal $(\mPhi,\mS)$  \\ 
&                     &            &       &         & $=(1000,150)\gev$ & $=(600,150)\gev$ & $=(400,100)\gev$  \\ 
&                     &            &       &         & $c\tau=1.17$~m & $c\tau=1.72$~m & $c\tau=1.46$~m \\
\hline
\textbf{Preselection:} & Pass trigger, 2 clean jets \& $\sumMinDR >0.5$  &  $  1375483 $  &  $ 183015 $  &  $526.0$ &   $ 26.2\% $  &     $ 22.4\% $  &     $ 17.5\% $  & \\
\textbf{Event cleaning:} & $\EvbdtLITau>0.1$  &  $ 4515 $  &     $ 192 $  &  $7.6$ &  $ 25.4\% $  &  $ 21.2\% $  &     $ 15.3\% $  \\
 & Trigger matching  &  $ 3627 $  &     $ 119 $ &  $3.8$ &  $ 24.5\% $  &     $ 20.4\% $  &     $ 15.0\% $ \\
  & $-3 < t < 15~\ns$  &   $ 3388 $  &     $ 110 $ &3.2 &    $ 24.0\% $  &       $ 20.0\% $  &       $ 14.8\% $\\
	\textbf{High-\et selection:} &$\sumLogR>1$  &  $ 1815 $  &     $ 61 $ &$2.7$   &    $ 21.7\% $  &     $ 16.8\% $  &     $ 11.5\% $\\
	 &$\mhtoht<0.6$  &    $ 1421 $  &     $ 41 $   & $2.1$ & $ 18.1\% $  &     $ 15.2\% $  &     $ 10.9\% $ \\
	 & $\FCRjetpt>160\GeV $  &    $ 774 $  &      $ 26 $ & $0$ & $ 17.5\% $  &     $ 13.6\% $  &     $ 7.50\% $\\
	 & $\SCRjetpt>100\GeV $  &    $ 459 $  &      $ 15 $ & $0$ &  $ 16.5\% $  &     $ 11.8\% $  &     $ 5.56\% $ \\
	 \textbf{Region A :}  &    & 10  &   $ 1 $        &   $0$  &  $ 10.7\% $  &     $ 7.74\% $  &     $ 3.10\% $\\
\hline
\hline

& Low-\et selection          &       Main data &   BIB & Cosmic rays & &Signal $(\mPhi,\mS)$  &  Signal $(\mPhi,\mS)$ \\
&                            &                 &       &   & &  $=(200,50)\gev $ &  $=(125,25)\gev$ \\
&                            &                 &       &   & &  $c\tau=1.07$~m   &  $c\tau=0.76$~m \\
\hline
 \textbf{Preselection:} & Pass trigger, 2 clean jets \& $\sumMinDR >0.5$  &   2180349  &   95247  & 319.1  & &   $ 7.58\% $  &     $ 4.33\% $ \\
 \textbf{Event cleaning:} & $\EvbdtLITopo>0.1$  &  $ 40474 $  &    $ 678 $  & 65.1 & &  $ 6.26\% $  &     $ 2.73\% $\\
  & Trigger matching  &  $ 34567 $  &    $ 538 $ &42.1 &&   $ 5.97\% $  &     $ 2.51\% $ \\
	 & $-3 < t < 15~\ns$  &   $ 33680 $  &    $ 519 $ & 23.4  & &  $ 5.86\% $  &     $ 2.46\% $ \\
	 \textbf{Low-\et selection:} &$\sumLogR>2.5$  &   $ 722 $  &      $ 13 $ & 18.3  & &   $ 0.92\% $  &    $ 0.39\% $\\
	 & $\FCRjetpt>80\GeV$  &   $ 304 $  &      $ 6 $ & 7.3  & &  $ 0.69\% $  &    $ 0.16\% $  \\
	 & $\SCRjetpt>60\GeV$  &    $ 136 $  &      $ 4 $ & 3.5  & &    $ 0.60\% $  &    $ 0.10\% $  \\
	 \textbf{Region A:}  &    & 7  &   0 & 0.4  & &  $ 0.43\% $  &    $ 0.07\% $  \\
	 \hline
\end{tabular}
}
\label{tab:Selection_cutFlow}
\end{table}
\end{landscape}

The validity of the ABCD method is tested by applying it to two validation regions (VRs).
These are similar to the main selections, but have modified requirements and boundaries for the ABCD plane variables, to ensure orthogonality to the high-\et and low-\et selections.
The VR for the high-\et selection (\VRITau) is defined as the nominal selection except for requiring $100 < \FCRjetpt < 160~\GeV$ and it is evaluated in the ABCD plane defined within $0.1<\evbdtLITau<0.22$.
The VR for the low-\et selection (\VRITopo) is defined as the nominal selection and it is evaluated in the ABCD plane defined within $0.1<\evbdtLITopo<0.22$.

In both VRs, the correlation observed between the two variables defining the ABCD plane is negligible ($< 3\%$ in main data) and signal contamination in region A is small. In all cases, the estimated number of background events is in good agreement with the number of data events observed in region A, as summarised in Table~\ref{table:ABCD_VR}. 

\begin{table}[ht]
\caption{Application of the ABCD method to the final high-\et and low-\et VRs. The column labelled ``Estim.~A'' gives the estimated contribution of background in the search region A assuming no signal, as calculated using the ABCD method. The statistical uncertainty of this calculation is also given. A, B, C and D show the number of observed events in each region. Only statistical uncertainties are considered in this table.
}
\begin{center}
\setlength{\tabcolsep}{0.0pc}
{\small
\begin{tabular*}{\textwidth}{@{\extracolsep{\fill}}lrrrrrrrrr}
Validation selections          &   Estim. A  &   A &     B &      C &       D \\
\hline
\VRITau  &  $66 \pm 15$ & 70 & 64 & 57 & 55 \\
\VRITopo  &  $54 \pm 17$ & 36 & 35 & 34 & 22 \\

 \hline
\end{tabular*}
}
\end{center}
\label{table:ABCD_VR}
\end{table}

The uncertainty in the data-driven background estimate is studied using a dijet-enriched sample.
This sample is selected using a single-jet-based trigger and vetoing on the CalRatio triggers to make sure that the event selection is orthogonal to the one used in the main analysis.
The ABCD planes are then defined similarly to those in the main analysis, but adjusting the boundaries in regions A, B, C and D to reduce the effect of statistical fluctuations in the estimation of the number of dijet events in region A given by the method.
The difference between the estimated and observed numbers of events in region A is taken as the systematic uncertainty associated with the method: $22\%$ in the high-\et ABCD plane and $25\%$ in the low-\et plane. The size of the statistical component of these uncertainties is $17\%$ and $20\%$, respectively.

The yields in each region of the main high-\et and low-\et selections are shown in the Table~\ref{table:ABCD_Selections} alongside the final background estimate calculated from a simultaneous background-only fit to all regions using the statistical model described in Section~\ref{sec:interpretation}. The expected background in each region is allowed to float so long as the ABCD relation is satisfied, with a Poisson constraint on the observed number of events in the corresponding region.
If the observed data in region A are ignored in the fit by removing the Poisson constraint on region A, the background estimate is the same as that expected from the ABCD relation ($N^\mathrm{bkg}_\mathrm{A}=(N^\mathrm{bkg}_\mathrm{B}\cdot N^\mathrm{bkg}_\mathrm{C})/N^\mathrm{bkg}_\mathrm{D}$), but with all sources of uncertainty accounted for. This corresponds to the \emph{a priori} (pre-unblinding) background estimate. The \emph{a posteriori} (post-unblinding) background estimate, which is used for the purposes of statistical interpretation, is obtained from the same background-only simultaneous fit to all regions, taking the observed number of events in A into account. Here also the ABCD relation is imposed, within the uncertainty of the ABCD method. When performing a signal-plus-background fit during the statistical interpretation, the estimated background can vary as a function of the signal strength.

\begin{table}[ht]
\caption{Application of the modified ABCD method to the final high-\et and low-\et selections. The columns A, B, C and D contain the number of observed events in each region.
``Estim.~A'' refers to the estimated contribution of background in the search region A assuming no signal, as calculated using the full statistical model decribed in Section~\ref{sec:interpretation}. The \emph{a priori} estimate refers to the ``pre-unblinding'' case, where the data in region A are ignored by removing the Poisson constraint in that region and the signal strength is fixed to zero.
This corresponds to the simple $N^\mathrm{bkg}_\mathrm{A}=(N^\mathrm{bkg}_\mathrm{B}\cdot N^\mathrm{bkg}_\mathrm{C})/N^\mathrm{bkg}_\mathrm{D}$ relation.
The \emph{a posteriori} estimate refers to the ``post-unblinding'' case, including the observed data in region A into the background only global fit, obtained by fixing the signal strength to 0. Only the fitted value in region A is shown, though the fitted number of events in regions B, C and D fluctuate as well. All experimental and statistical uncertainties are included in the quoted error.
}
\begin{center}
\setlength{\tabcolsep}{0.0pc}
{\small
\def\arraystretch{1.2}
\begin{tabular*}{\textwidth}{@{\extracolsep{\fill}}lrrrrrrrrrr}
Main selections         &     B &      C &     D &  Estim. A           &   A  &  Estim. A   \\
                        &       &        &       &  (\emph{a priori})  &      &  (\emph{a posteriori})          \\
\hline
High-\et selection      &   9   &   187  &  253  & $6.7^{+3.2}_{-2.3}$   &  10  & $8.5^{+2.3}_{-2.0} $ \\
Low-\et selection       &   2   &   70   &   57  & $2.5^{+2.5}_{-1.4}$   &   7  & $5.3^{+2.1}_{-1.6} $ \\

 \hline
\end{tabular*}
}
\end{center}
\label{table:ABCD_Selections}
\end{table}

\section{Systematic uncertainties}
\label{sec:systematics}

The uncertainty in the data-driven ABCD method for the background estimate is discussed in Section~\ref{sec:background}, and found to be $22\%$ in the high-\et ABCD plane and $25\%$ in the low-\et plane.

Several uncertainties related to modelling, theory and reconstruction affect the estimated signal yield.
The jet-energy scale and jet-energy resolution introduce uncertainties in the signal yield of 1\% to 9\% and 1\% to 5\%, respectively, depending on the model, where the high-\mPhi models are least affected. These uncertainties are calculated using the procedure detailed in Ref.~\cite{Aaboud:2017jcu}.
Since the jets used in this analysis are required to have a low fraction of calorimeter energy in the ECal, the jet-energy uncertainties are re-derived as a function of ECal energy fraction as well as of $\eta$.
The additional jet-energy uncertainties are found to have an effect of up to 17\% on the signal yield, and are conservatively taken in quadrature with the regular jet-energy uncertainties.
The lower-\mPhi models are more sensitive to all jet-energy uncertainties than the higher-\mPhi models.

The uncertainty in the signal trigger efficiency is estimated by studying how well modelled the three main HLT variables (jet \et and \lograt, and \pt of tracks within the jet) are between HLT- and offline-reconstructed quantities in data and Monte Carlo (MC) simulation. A tag-and-probe technique using standard jet triggers is used to obtain a pure sample of multijet events in both data and MC simulation. Scale factors are derived that represent the degree of mis-modelling in each variable, and are applied in an emulation of the CalRatio triggers. The change in yield relative to the nominal (unscaled) trigger emulation after the full analysis selection is taken as the size of the systematic uncertainty, which is 2\% or less for all models.

Events in MC simulation are reweighted to obtain the correct pileup distribution. A variation in the pileup reweighting of MC is included to cover the uncertainty on the ratio between the predicted and measured inelastic cross-section in the fiducial volume defined by $M_X > 13$ GeV where $M_X$ is the mass of the hadronic system~\cite{STDM-2015-05}.
The uncertainty in the pile-up reweighting of the reconstructed events in the MC simulation is estimated by comparing the distribution of the number of primary vertices in the MC simulation with the one in data as a function of the instantaneous luminosity. 
Differences between these distributions are adjusted by scaling the mean number of $pp$ interactions per bunch crossing in the MC simulation and the $\pm 1 \sigma$ uncertainties are assigned to these scaling factors. The effect on the signal event yields varies between
1\% and 12\% depending on the model. The low-\mPhi models are the most affected by this uncertainty.

The \textsc{NNPDF2.3LO}~\cite{Ball:2012cx} PDF set was used when generating the signal samples. In addition to the nominal PDF, 100 PDF variations are also included in the set. The PDF uncertainty is evaluated by taking the standard deviation of signal event yield when each of these PDF variations is used instead of the nominal.
The effect on the signal yield is between 3\% and 8\% depending on the signal sample, where the size of the uncertainty grows with \mPhi.

A systematic uncertainty is included to account for potential mis-modelling of BDT input variables, using the same control sample of dijet events defined for the evaluation of the systematic uncertainty in the data-driven background estimate.
In this control sample, the distributions of the inputs and outputs of the per-jet and per-event BDTs were studied, and were found to agree fairly well between data and MC simulation.
The residual differences are translated into a systematic uncertainty in the signal efficiency by randomly varying the input variables according to their uncertainty and re-evaluating the BDTs for each signal event. The value of the resulting uncertainty is up to 2\% depending on the model, where the largest uncertainties are assigned to the lower-\mPhi models.

Finally, the uncertainty in the integrated luminosity is around $2\%$. 
It is derived, following a methodology similar to that detailed in Ref.~\cite{DAPR-2013-01}, and using the LUCID-2 detector for the baseline luminosity measurements \cite{LUCID2}, from calibration of the luminosity scale using $x$-$y$ beam-separation scans.
This uncertainty affects all models equally.

\section{Statistical interpretation}
\label{sec:interpretation}

\subsection{Extraction of limits}
\label{sec:extractionOfLimits}

A data-driven background estimation and signal hypothesis test is performed simultaneously in all regions.
An overall profile likelihood function is constructed from the product of the Poisson probabilities of observing the number of events $N^\textrm{obs}_{X}$, given an expectation $N^\textrm{exp}_{X}$, in each region $X$, where $X = {\textrm{A},\textrm{B}, \textrm{C}, \textrm{D}}$.
The value of $N^\textrm{exp}_{X}$ in each region is the sum of: the expected signal yield $N^\textrm{sig}_{X}$, given by the number of simulated signal events entering region $X$ multiplied by the signal strength $\mu$ (the parameter of interest); and the expected background yield $N^\textrm{bkg}_{X}$.
In the fit, the expected background yields are constrained to obey the ABCD relation $N^\textrm{bkg}_\textrm{A}=(N^\textrm{bkg}_\textrm{B}\cdot N^\textrm{bkg}_\textrm{C})/N^\textrm{bkg}_\textrm{D}$. This reduces the number of degrees of freedom of the fit by one as $N^\textrm{obs}_\textrm{A} = m N^\textrm{bkg}_\textrm{B} + \mu N^\textrm{sig}_\textrm{A}$ and $N^\textrm{obs}_\textrm{C} = m N^\textrm{bkg}_\textrm{D} + \mu N^\textrm{sig}_\textrm{C}$, where $m$ is a free parameter.
Since the Poisson constraints only apply to $N^\textrm{obs}_{X}$ relative to $N^\textrm{exp}_{X}$, it follows that the background prediction may change dynamically in the fit as a function of the signal strength.

As can be seen in Table~\ref{table:ABCD_Selections}, no excess of events is observed in region A for either of the analysis selections. The CL$_\textrm{s}$ method~\cite{Read:2002hq} is therefore used to set upper limits on $\sigma(\Phi) \times \BR $ in the benchmark HS model.

Systematic uncertainties for signal, background and luminosity are represented by nuisance parameters. Each nuisance parameter is assigned a Gaussian constraint of relevant width (see Section~\ref{sec:systematics}).
An asymptotic approach~\cite{Cowan:2010js} is used to compute the CL$_\textrm{s}$ value, and the limits are defined by the region excluded at 95\% confidence level (CL). The asymptotic approximation was tested and found to give consistent results with limits obtained from ensemble tests.

Since each signal sample was generated for a particular LLP proper decay length, it is necessary to extrapolate the signal efficiency to other decay lengths to obtain limits as a function of $c\tau$. This is achieved by using a weighting method, which is applied separately to each signal sample. The weight to be assigned to a displaced jet with lifetime $\tau_{\textrm{new}}$ is obtained from the sample generated with lifetime $\tau_{\textrm{gen}}$ by:

\begin{equation*}
  w(t) = \frac{\tau_\textrm{gen}}{\exp(-t/\tau_\textrm{gen})} \cdot \frac{\exp(-t/\tau_\textrm{new})}{\tau_\textrm{new}}.
\end{equation*}

The quantity $t$ is the proper decay time of the LLP that gives rise to the displaced jet. In the benchmark HS model, the LLPs are pair produced, so each event is weighted by the product of the individual LLP weights. The weighted sample is used to evaluate the signal efficiency for $c\tau_{\textrm{new}}$.

\begin{figure}[ht]
\centering
\includegraphics[width=0.497\linewidth]{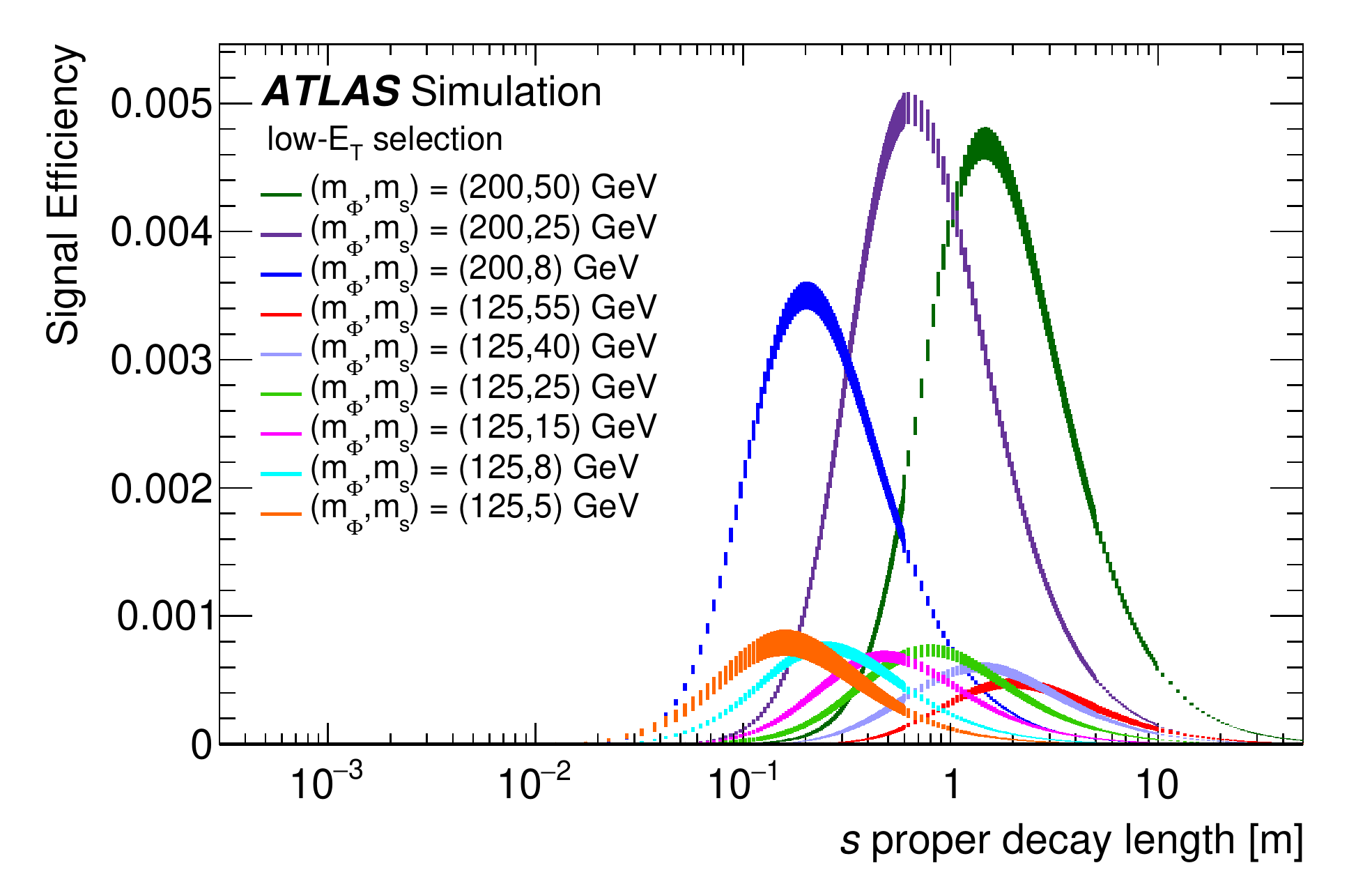}
\includegraphics[width=0.497\linewidth]{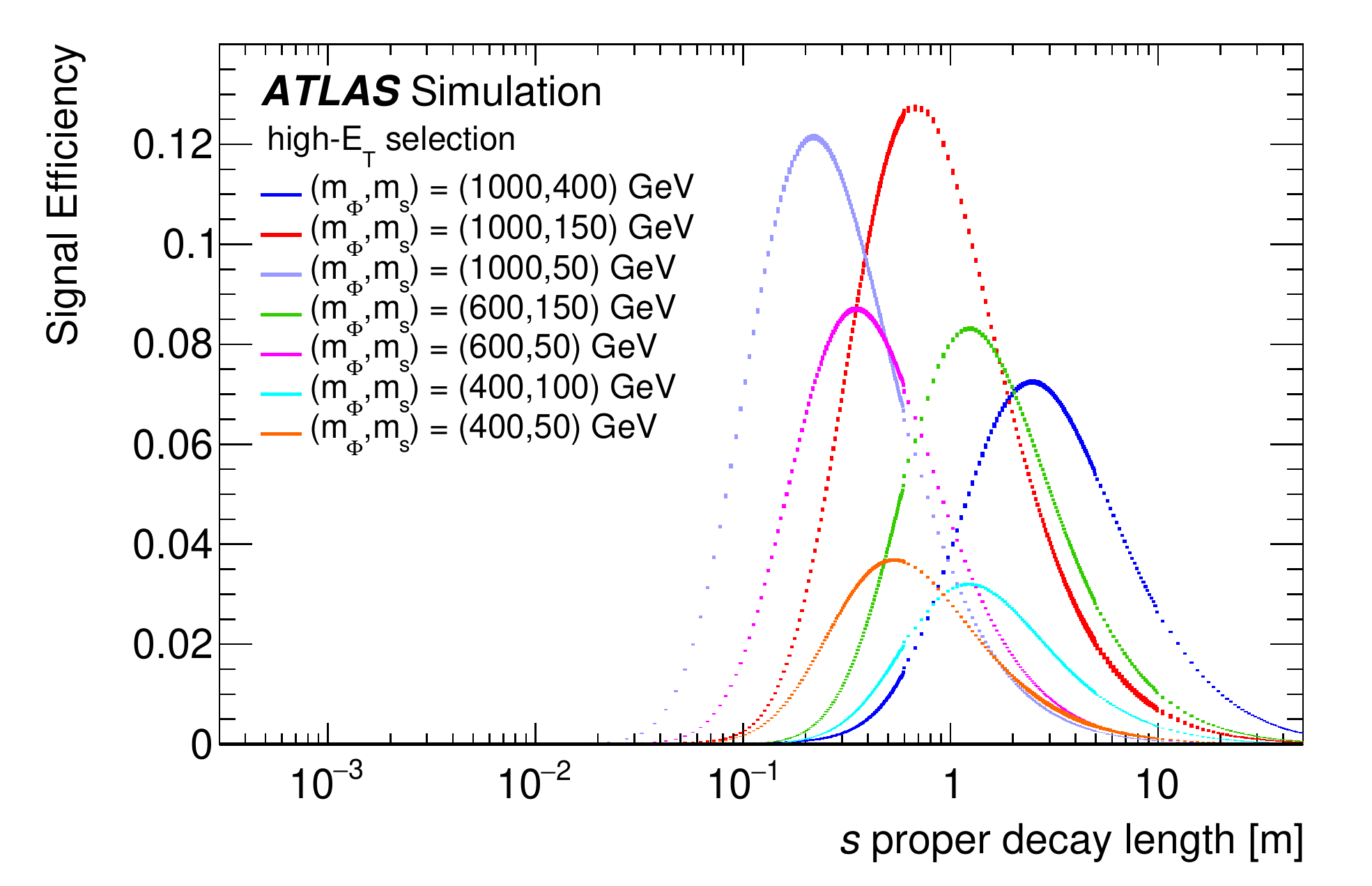}
\caption{The extrapolated signal efficiencies as a function of proper decay length of the \textit{s} for several simulated samples in the low-\et (left) and high-\et (right) selections. The vertical error bars represent the statistical uncertainties.}
\label{fig:sig-eff-vs-lifetime}
\end{figure}

The upper limit at a given $c\tau$ is then obtained by scaling the limit at $c\tau_{\textrm{gen}}$ by the ratio of signal efficiencies at $c\tau$ and $c\tau_{\textrm{gen}}$. This procedure for extrapolating the efficiency to different lifetimes was checked by comparing the extrapolated efficiency derived from the main simulated samples with the measured efficiency of samples with alternative LLP lifetime assumptions. These were found to agree within statistical uncertainties. Figure~\ref{fig:sig-eff-vs-lifetime} shows the extrapolated efficiency for the signal samples with \mPhi of 125 and $200 \GeV$ with the low-\et selection applied, alongside the efficiency for signal samples with \mPhi of $400 \GeV$, $600 \GeV$, and $1 \TeV$ signal samples with the high-\et selection applied.

\begin{figure}[ht]
\centering
\includegraphics[width=0.68\linewidth]{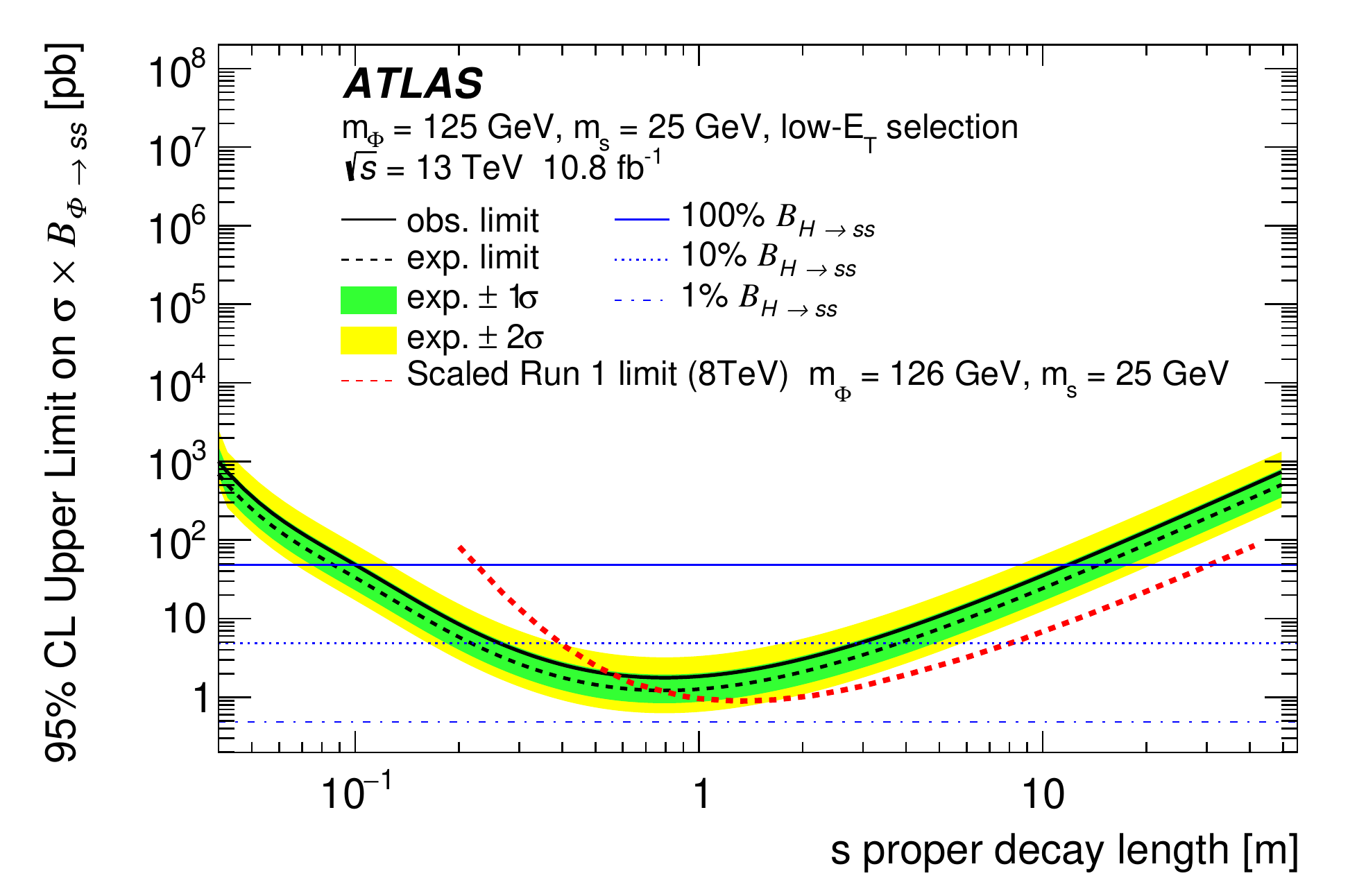}\\
\includegraphics[width=0.68\linewidth]{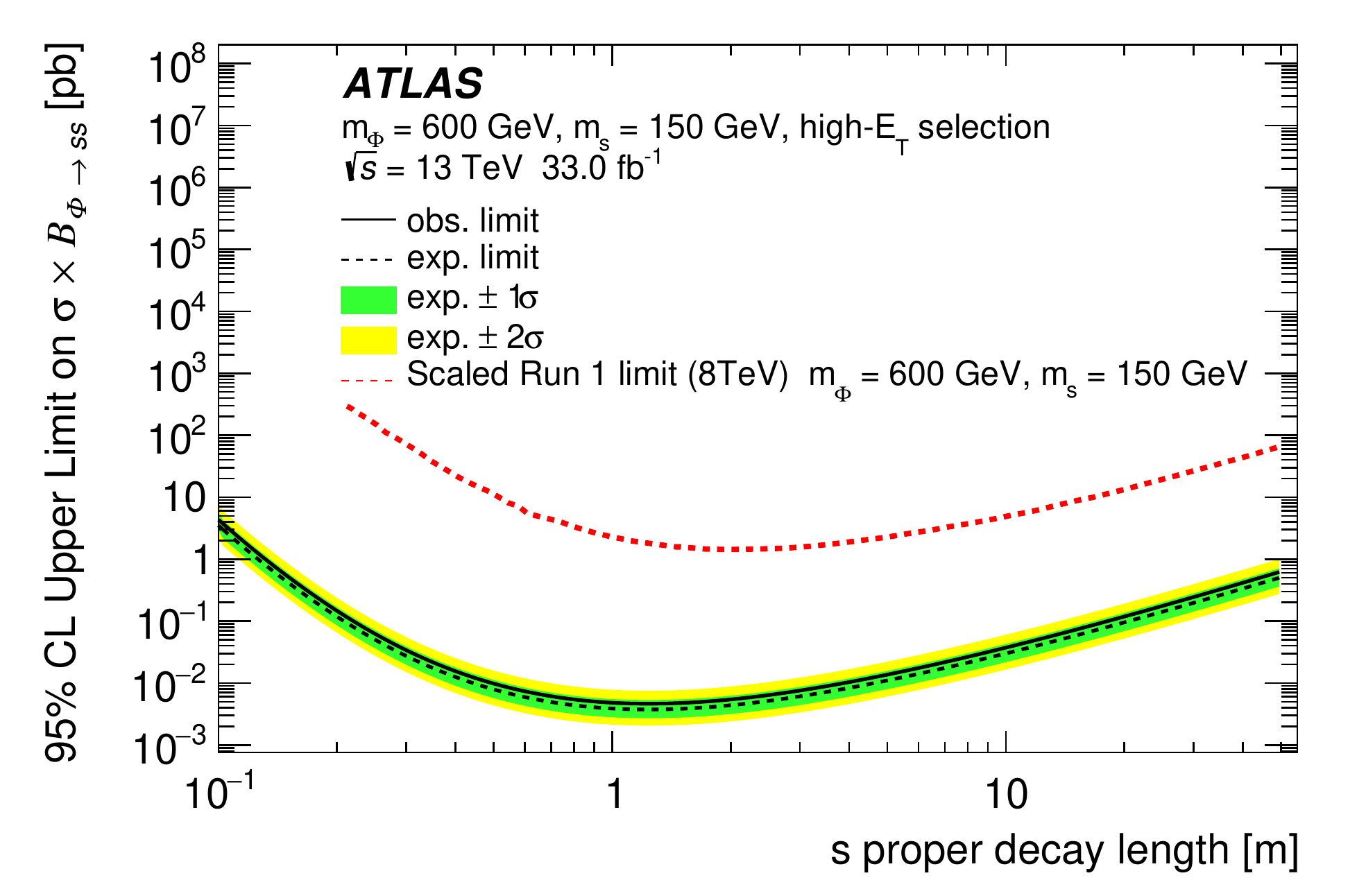}\\
\caption{The observed limits, expected limits and $\pm 1\sigma$ and $\pm 2\sigma$ bands for two models with $\mPhi=125\gev,\mS=25\gev$ and $\mPhi=600\gev,\mS=150\gev$. The top plot also shows the SM Higgs boson gluon--gluon fusion production cross section for $m_{H}=125\GeV$, assumed to be 48.58~pb at 13~\TeV~\cite{LHCHXSWGYR4}. Both plots show a comparison with the limits obtained for a comparable model in the Run 1 analysis~\cite{EXOT-2012-28} scaled by the ratio of parton luminosities for gluon--gluon fusion between $13\TeV$ and $8\TeV$ for a particle of appropriate mass.}
\label{fig:example-brazil-brands}
\end{figure}

\begin{figure}[ht]
\centering
\includegraphics[width=0.495\linewidth]{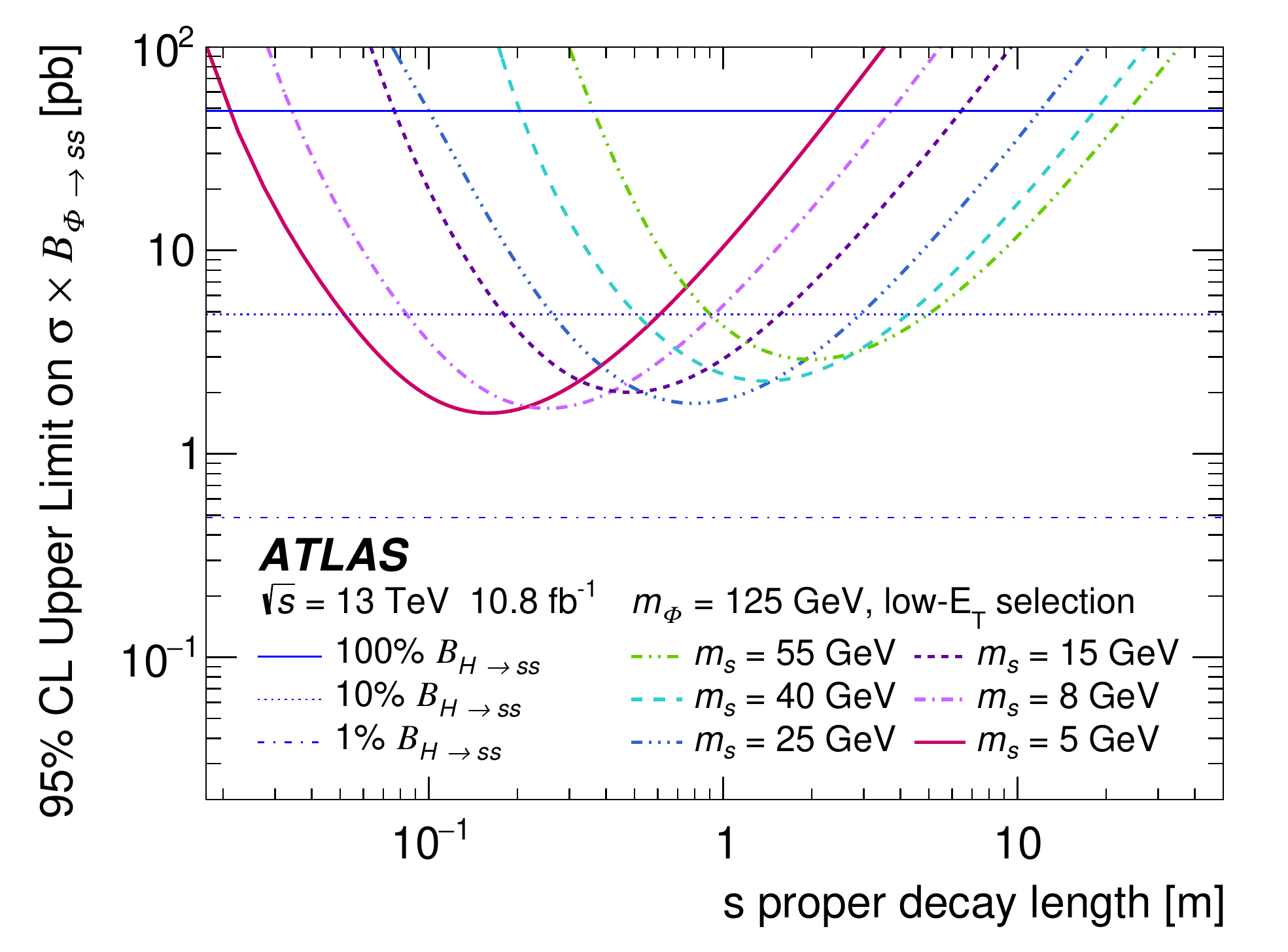}
\includegraphics[width=0.495\linewidth]{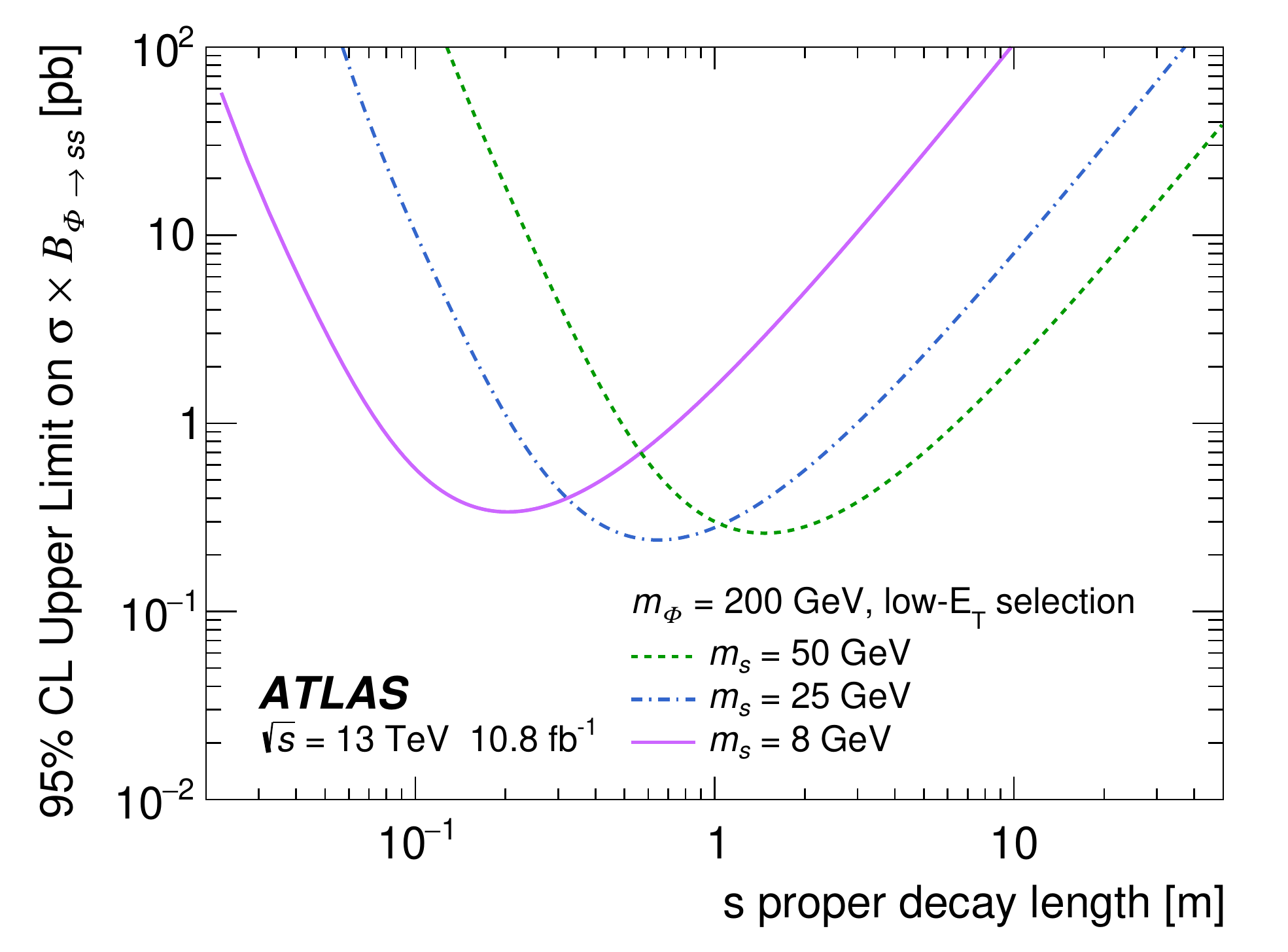}
\includegraphics[width=0.495\linewidth]{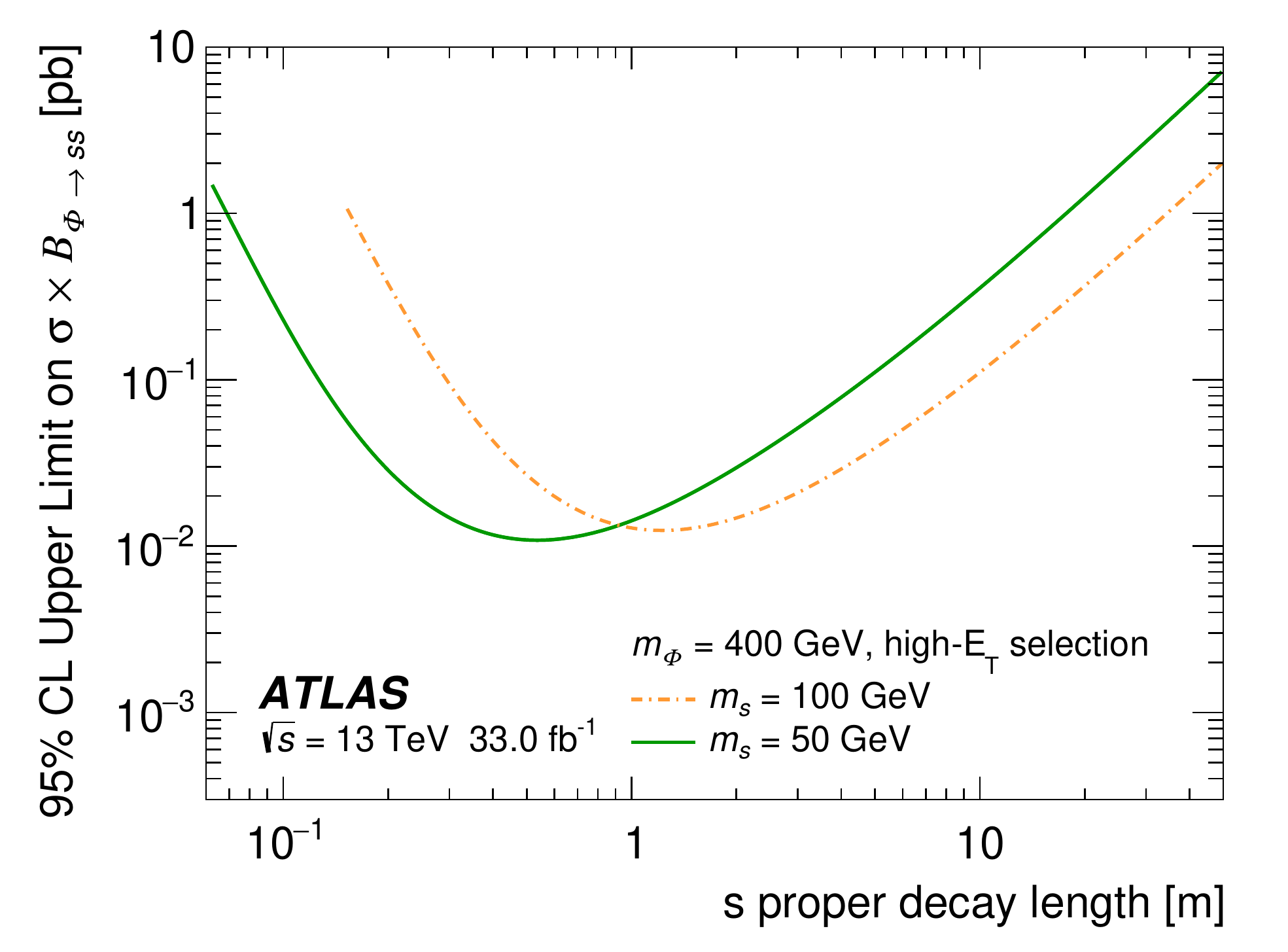}
\includegraphics[width=0.495\linewidth]{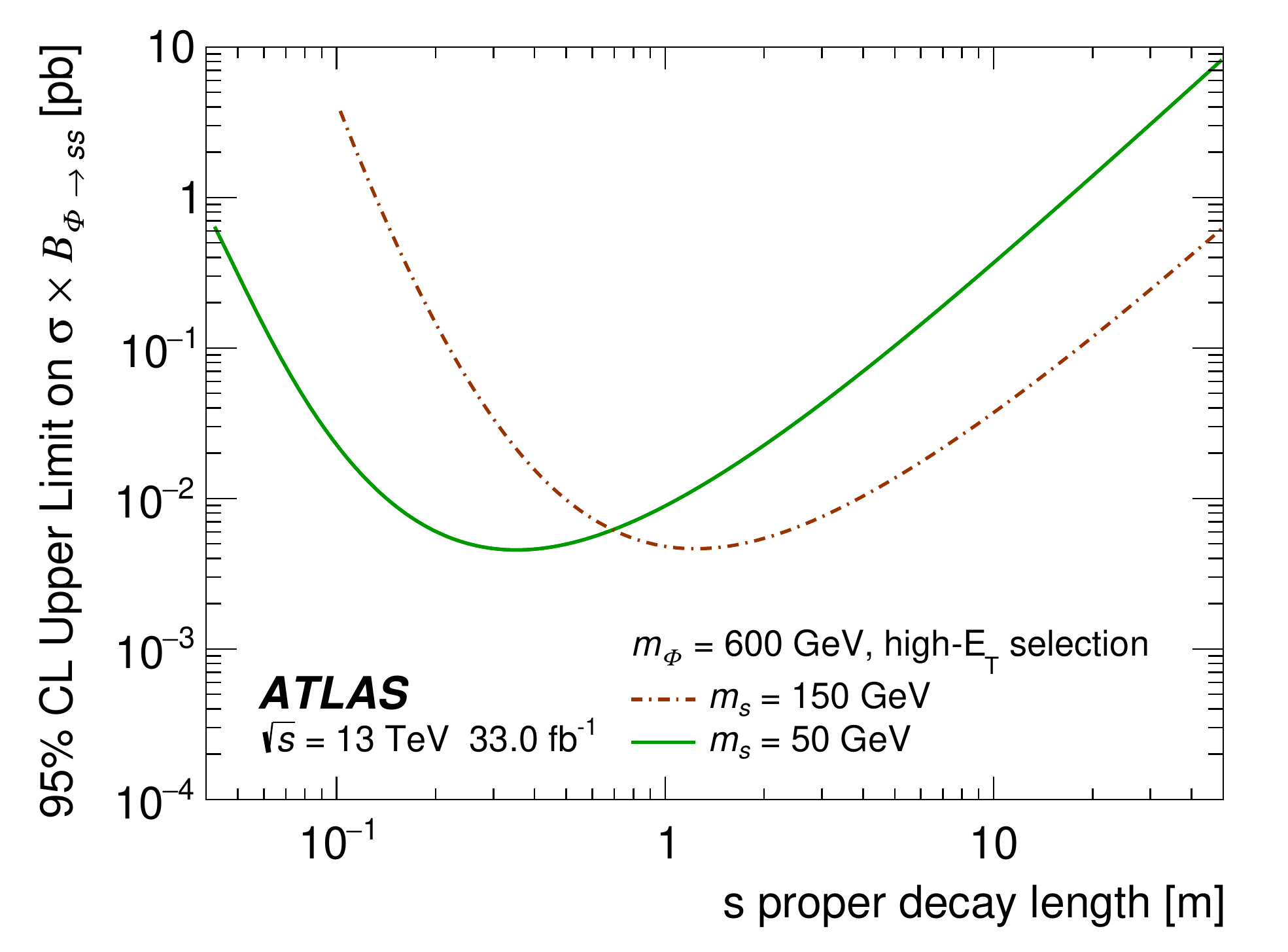}
\includegraphics[width=0.495\linewidth]{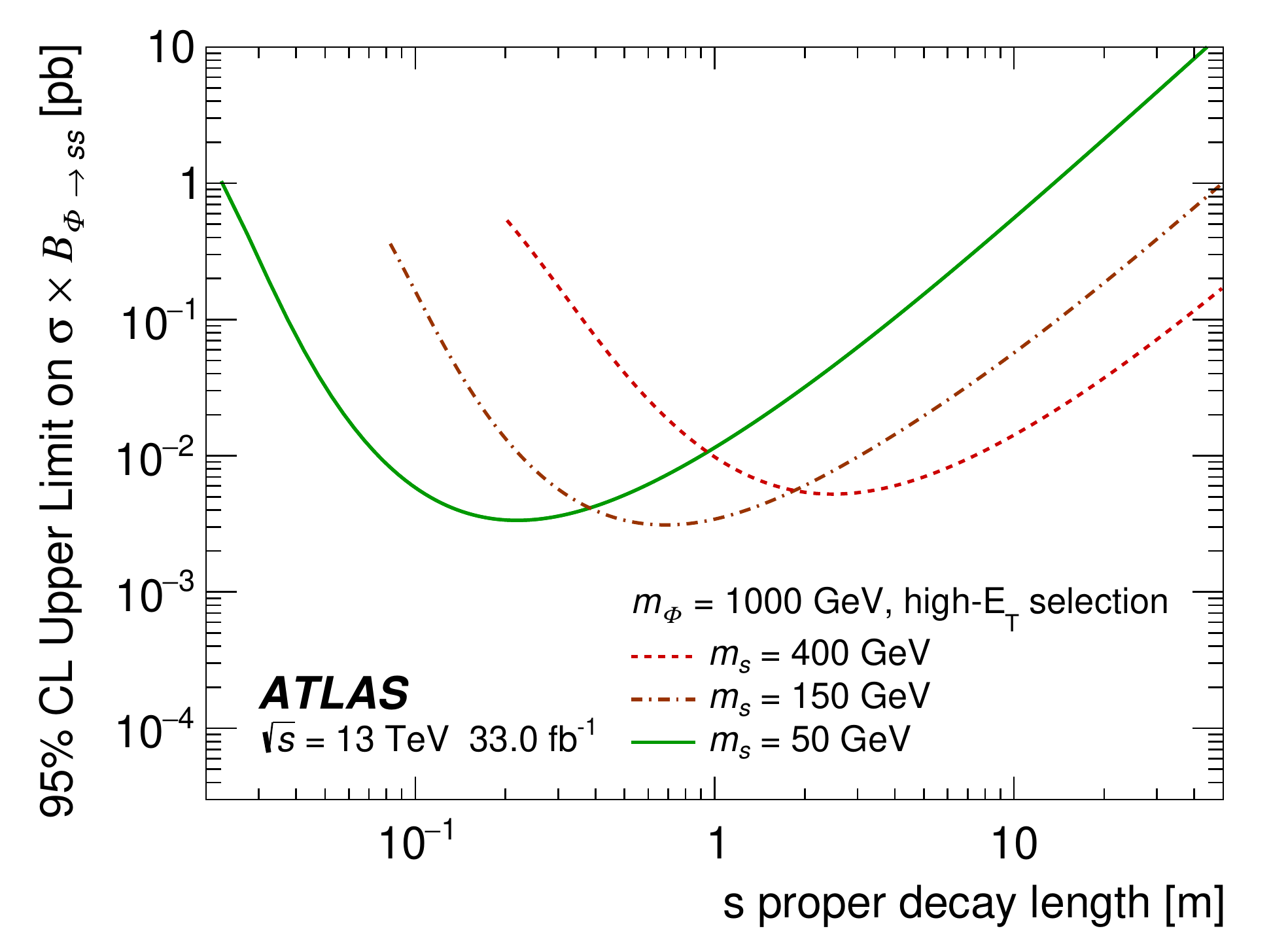}
\caption{The observed limits for the $\Phi$ masses of 125, 200, 400, 600 and $1000\GeV$.}
\label{fig:limits}
\end{figure}

The observed and expected limits for two example signal models can be seen in Figure~\ref{fig:example-brazil-brands}. The observed limits for all considered models are summarised in Figure~\ref{fig:limits}. The expected limits correspond to those obtained using the \emph{a posteriori} background estimate, which is given in Table~\ref{table:ABCD_Selections}. This explains why the observed and expected limits may appear closer than anticipated from the observed and expected numbers of events in region A using the simple ABCD relation.

For a mediator similar to the Higgs boson and of mass $\mPhi = 125\GeV$, the limits are presented divided by the SM Higgs boson gluon--gluon fusion production cross section for $m_{H}=125\GeV$, assumed to be 48.58~pb at 13~\TeV~\cite{LHCHXSWGYR4}. For such models, decays of neutral scalars with masses between $5$ and $55~\GeV$ are excluded for proper decay lengths between 5~cm and 5~m depending on the LLP mass (assuming a 10\% branching ratio). 
Compared with the 8~\TeV results, the limits for models with $\mPhi= 125\GeV$ are typically a factor 10 more stringent around 20~cm and a factor 10 less stringent around 50~m.

For $\mPhi = 200~\GeV$, cross section times branching ratio values above 1~pb are ruled out between 5~cm and 7~m depending on the scalar mass. 
For models with $\mPhi = 400\GeV$, $\mPhi = 600\GeV$, and $\mPhi = 1000\GeV$,  $\sigma(\Phi) \times \BR$ values above 0.1~pb are ruled out at 95\% CL between about 12~cm and 9~m, 7~cm and 20~m, and 4~cm and 35~m respectively, depending on the scalar masses. 
The limits are significantly more stringent than the 8~\TeV results across the whole lifetime range, and in some cases limits are set on combinations of \mPhi and \mS that were not previously studied.

\subsection{Combination of results with MS displaced jets search}

In this section the limits derived in Section~\ref{sec:extractionOfLimits} are combined with the results for the comparable models from the muon spectrometer (MS) displaced-jets analysis~\cite{Aaboud:2018aqj}. 
The MS analysis searches for neutral LLPs decaying at the outer edge of the HCal or in the MS. 
These decays result in secondary-decay vertices that can be reconstructed as displaced vertices in the MS.
The analysis considers events containing either two displaced vertices in the MS or one displaced vertex together with prompt jets or \met.
Some of the benchmark models used in the MS vertex search are the same models considered in the search described in this paper.
Therefore a combination of the results of these two complementary analyses can be performed.

The orthogonality of the CalRatio (CR) and MS analyses was checked in both data and simulated signal to ensure the final selections were statistically independent.
The combination is performed using a simultaneous fit of the likelihood functions of each analysis. 
The signal strength as well as the nuisance parameter for the luminosity uncertainties is chosen to be the same for the CR and MS likelihoods. The  signal uncertainties are chosen to be uncorrelated, since they are dominated by different experimental uncertainties in the two searches. The effect of correlating the signal uncertainties was studied by comparing the limit obtained with no correlation in signal uncertainties to that obtained with correlation of relevant signal uncertainties. The effect on the combined limits was found to be negligible. The background estimate in each analysis is data-driven and the two estimates are therefore not correlated.

As in the individual searches, the asymptotic approach is used to compute the CL$_\textrm{s}$ value, and the limits are defined by the region excluded at 95\% CL.
The limits are calculated using a global fit, where the overall likelihood function is the product of the individual likelihood functions of the searches to be combined. 
The limits are calculated separately at each point in the $c\tau$ range of interest, where in each case the signal efficiency is scaled by the result of the lifetime extrapolation. 

\begin{figure}[ht]
\centering
 \includegraphics[width=0.68\linewidth]{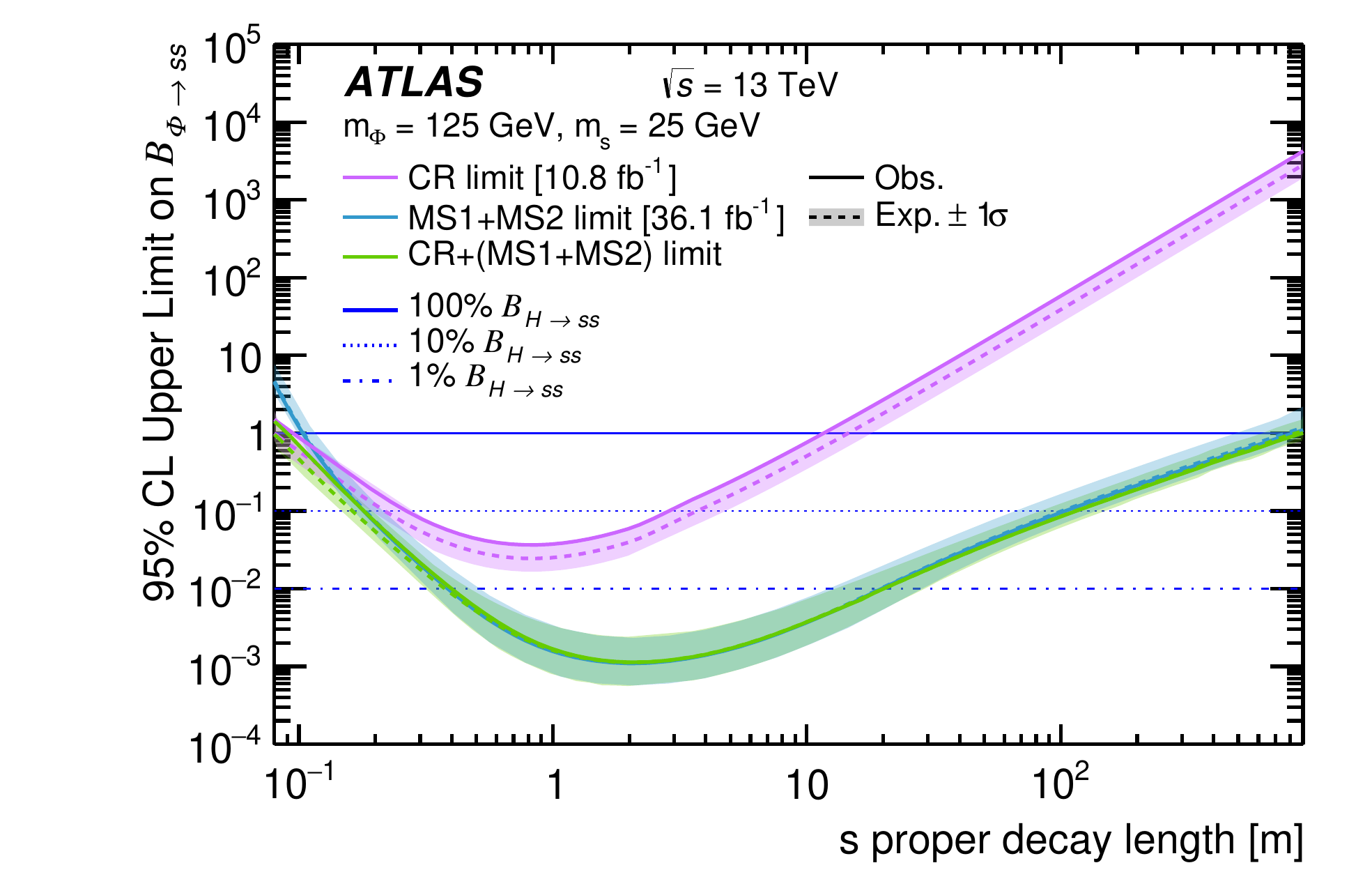}\\
 \includegraphics[width=0.68\linewidth]{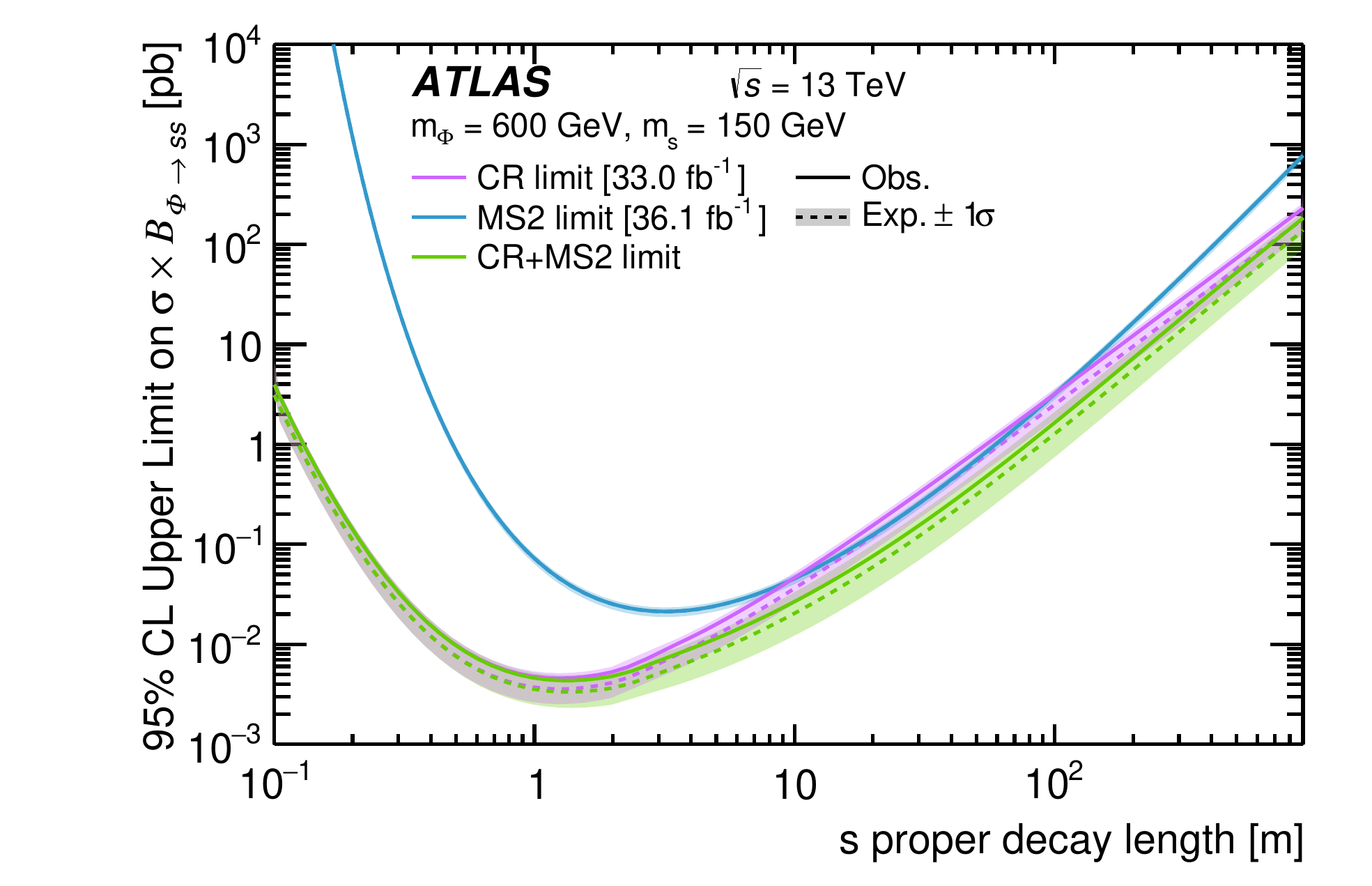}
\caption{Examples of the combined limits for models with $\mPhi=125\GeV$ and $\mPhi=600\GeV$ from the CR analysis and the MS analysis, which is separated into the MS 1-vertex plus \met (MS1) and MS 2-vertex (MS2) components. The MS1 component of the MS displaced jet search was only applied to models with $\mPhi=125\GeV$. The expected limit is shown as a dashed line with shading for the $\pm 1\sigma$ band, while the observed is a solid line. The colours of the shading and solid and dashed lines refer to the limits from each analysis and their combination, as indicated in the legend.
}
\label{fig:comb}
\end{figure}

The observed and expected limits for two example signal models are shown in Figure~\ref{fig:comb}. 
For the models with $\mPhi = 125~\GeV$, the MS analysis has higher sensitivity than the CR analysis at large decay lengths. 
For short decay lengths ($<10$~cm) the sensitivities of the two analyses are comparable and the combination of their limits provides a slight improvement.
The limits for intermediate masses, $\mPhi = 200~\GeV$ and $400~\GeV$, show a clear complementarity of the analyses: the CR limits, which improve with $\mPhi$, are stronger at shorter decay lengths, while the MS analysis sets stronger limits at large decay lengths.
In this case the combination of the two analyses improves on the individual limits over the full range of decay lengths.
For higher masses, $\mPhi \geq 600~\GeV$, the CR analysis is in general more sensitive than the MS analysis. Even in this case, the combination provides a modest improvement on the CR-only limit at long decay lengths.

\FloatBarrier

\section{Conclusion}
\label{sec:conclusion}

A search for pair-produced long-lived particles decaying in the ATLAS calorimeter is presented, using data collected during $pp$ collisions at the LHC in 2016, at centre-of-mass energy of 13~\TeV. The dataset size is  \intlumiNOMATCH or \intlumi depending on whether the data were collected using a low- or high-\ET dedicated trigger. Benchmark hidden-sector models are used to set limits, where the mediator's mass ranges between 125 and 1000~\GeV, while the long-lived scalar's mass range between 5 and 400~\GeV.
The search selects events with two signal-like jets (which are typically narrow, trackless, and with a large fraction of their energy in the hadronic calorimeter) using machine-learning techniques. Two signal regions are defined for the low- and high-\ET datasets. The background estimation is performed using the data-driven ABCD method.
No significant excess is observed in either signal region. The CL$_\textrm{s}$ method is therefore used to set 95\% CL limits on $\sigma(\Phi) \times \BR $ as a function of LLP decay length. 
For a mediator similar to the Higgs boson and of mass $\mPhi = 125\GeV$, decays of neutral scalars with masses between $5$ and $55~\GeV$ are excluded for proper decay lengths between 5~cm and 5~m depending on the LLP mass (assuming a 10\% branching ratio). 
For $\mPhi = 200~\GeV$, cross section times branching ratio values above 1~pb are ruled out between 5~cm and 7~m depending on the scalar mass. 
For models with $\mPhi = 400\GeV$, $\mPhi = 600\GeV$, and $\mPhi = 1000\GeV$,  $\sigma(\Phi) \times \BR$ values above 0.1~pb are ruled out between about 12~cm and 9~m, 7~cm and 20~m, and 4~cm and 35~m respectively, depending on the scalar masses. 
A combination of the limits with the results of a similar ATLAS search looking for displaced vertices in the muon spectrometer is performed.
The resulting combined limits provide a summary of the ATLAS results for pair-produced neutral LLPs. The combined limits tend to follow the results from the most sensitive search for each mediator: for low mediator masses ($\mPhi\leq 200\gev$), the sensitivity is dominated at high decay lengths by the muon spectrometer limits and at very low decays lengths by the CalRatio limits. For higher mediator masses ($\mPhi > 200\gev$), the sensitivity is dominated by the CalRatio search across most of the range of considered decay lengths. A small improvement in the overall limits is observed in regions where the two analyses have similar sensitivity.

\FloatBarrier

\section*{Acknowledgements}


We thank CERN for the very successful operation of the LHC, as well as the
support staff from our institutions without whom ATLAS could not be
operated efficiently.

We acknowledge the support of ANPCyT, Argentina; YerPhI, Armenia; ARC, Australia; BMWFW and FWF, Austria; ANAS, Azerbaijan; SSTC, Belarus; CNPq and FAPESP, Brazil; NSERC, NRC and CFI, Canada; CERN; CONICYT, Chile; CAS, MOST and NSFC, China; COLCIENCIAS, Colombia; MSMT CR, MPO CR and VSC CR, Czech Republic; DNRF and DNSRC, Denmark; IN2P3-CNRS, CEA-DRF/IRFU, France; SRNSFG, Georgia; BMBF, HGF, and MPG, Germany; GSRT, Greece; RGC, Hong Kong SAR, China; ISF and Benoziyo Center, Israel; INFN, Italy; MEXT and JSPS, Japan; CNRST, Morocco; NWO, Netherlands; RCN, Norway; MNiSW and NCN, Poland; FCT, Portugal; MNE/IFA, Romania; MES of Russia and NRC KI, Russian Federation; JINR; MESTD, Serbia; MSSR, Slovakia; ARRS and MIZ\v{S}, Slovenia; DST/NRF, South Africa; MINECO, Spain; SRC and Wallenberg Foundation, Sweden; SERI, SNSF and Cantons of Bern and Geneva, Switzerland; MOST, Taiwan; TAEK, Turkey; STFC, United Kingdom; DOE and NSF, United States of America. In addition, individual groups and members have received support from BCKDF, CANARIE, CRC and Compute Canada, Canada; COST, ERC, ERDF, Horizon 2020, and Marie Sk{\l}odowska-Curie Actions, European Union; Investissements d' Avenir Labex and Idex, ANR, France; DFG and AvH Foundation, Germany; Herakleitos, Thales and Aristeia programmes co-financed by EU-ESF and the Greek NSRF, Greece; BSF-NSF and GIF, Israel; CERCA Programme Generalitat de Catalunya, Spain; The Royal Society and Leverhulme Trust, United Kingdom. 

The crucial computing support from all WLCG partners is acknowledged gratefully, in particular from CERN, the ATLAS Tier-1 facilities at TRIUMF (Canada), NDGF (Denmark, Norway, Sweden), CC-IN2P3 (France), KIT/GridKA (Germany), INFN-CNAF (Italy), NL-T1 (Netherlands), PIC (Spain), ASGC (Taiwan), RAL (UK) and BNL (USA), the Tier-2 facilities worldwide and large non-WLCG resource providers. Major contributors of computing resources are listed in Ref.~\cite{ATL-GEN-PUB-2016-002}.

\clearpage
\printbibliography
\clearpage 
 
\begin{flushleft}
{\Large The ATLAS Collaboration}

\bigskip

M.~Aaboud$^\textrm{\scriptsize 35d}$,    
G.~Aad$^\textrm{\scriptsize 101}$,    
B.~Abbott$^\textrm{\scriptsize 128}$,    
D.C.~Abbott$^\textrm{\scriptsize 102}$,    
O.~Abdinov$^\textrm{\scriptsize 13,*}$,    
D.K.~Abhayasinghe$^\textrm{\scriptsize 93}$,    
S.H.~Abidi$^\textrm{\scriptsize 167}$,    
O.S.~AbouZeid$^\textrm{\scriptsize 40}$,    
N.L.~Abraham$^\textrm{\scriptsize 156}$,    
H.~Abramowicz$^\textrm{\scriptsize 161}$
% ...
\bigskip
\\

% ...
$^{13}$Institute of Physics, Azerbaijan Academy of Sciences, Baku; Azerbaijan.\\
% ...
$^{35}$$^{(a)}$Facult\'e des Sciences Ain Chock, R\'eseau Universitaire de Physique des Hautes Energies - Universit\'e Hassan II, Casablanca;$^{(b)}$Facult\'{e} des Sciences, Universit\'{e} Ibn-Tofail, K\'{e}nitra;$^{(c)}$Facult\'e des Sciences Semlalia, Universit\'e Cadi Ayyad, LPHEA-Marrakech;$^{(d)}$Facult\'e des Sciences, Universit\'e Mohamed Premier and LPTPM, Oujda;$^{(e)}$Facult\'e des sciences, Universit\'e Mohammed V, Rabat; Morocco.\\
% ...
$^{40}$Niels Bohr Institute, University of Copenhagen, Copenhagen; Denmark.\\
% ...
$^{93}$Department of Physics, Royal Holloway University of London, Egham; United Kingdom.\\
% ...
$^{101}$CPPM, Aix-Marseille Universit\'e, CNRS/IN2P3, Marseille; France.\\
$^{102}$Department of Physics, University of Massachusetts, Amherst MA; United States of America.\\
% ...
$^{128}$Homer L. Dodge Department of Physics and Astronomy, University of Oklahoma, Norman OK; United States of America.\\
% ...
$^{156}$Department of Physics and Astronomy, University of Sussex, Brighton; United Kingdom.\\
% ...
$^{161}$Raymond and Beverly Sackler School of Physics and Astronomy, Tel Aviv University, Tel Aviv; Israel.\\
% ...
$^{167}$Department of Physics, University of Toronto, Toronto ON; Canada.\\
% ...
$^{a}$ Also at Centre for High Performance Computing, CSIR Campus, Rosebank, Cape Town; South Africa.\\
$^{b}$ Also at CERN, Geneva; Switzerland.\\
$^{c}$ Also at CPPM, Aix-Marseille Universit\'e, CNRS/IN2P3, Marseille; France.\\
$^{d}$ Also at D\'epartement de Physique Nucl\'eaire et Corpusculaire, Universit\'e de Gen\`eve, Gen\`eve; Switzerland.\\
$^{e}$ Also at Departament de Fisica de la Universitat Autonoma de Barcelona, Barcelona; Spain.\\
% ...
$^{*}$ Deceased

\end{flushleft}


\end{document}